\documentclass[12pt]{article}

\usepackage[utf8]{inputenc}
\usepackage[margin = 1.0in]{geometry}
\usepackage{amsmath}
\numberwithin{equation}{section}
\numberwithin{figure}{section}
\numberwithin{table}{section}
\usepackage{amsthm}
\usepackage{amssymb}
\usepackage{physics}
\usepackage{textcomp}
\usepackage{gensymb}
\usepackage{bbm}
\usepackage{graphicx}
\usepackage[dvipsnames]{xcolor}
\usepackage{quiver}
\usepackage{bbm}
\usepackage{bm}
\usepackage{float}
\usepackage{mathtools}
\usepackage{cases}
\usepackage[affil-it]{authblk}
\usepackage{hyperref}
\hypersetup{
    colorlinks,
    citecolor=blue,
    filecolor=black,
    linkcolor=Blue,
    urlcolor=blue,
    linktoc=all
}
\usepackage{tikz}
\usetikzlibrary{knots, decorations.markings, decorations.shapes, hobby}
\usetikzlibrary{arrows, calc}
\usepackage[justification=centering]{caption}
\usepackage{subcaption}
\usepackage{diagbox}
\usepackage{cellspace} 
\usepackage[none]{hyphenat}
\usepackage{cite}

\usepackage[scr=dutchcal]{mathalpha}
\newcommand{\scriptr}{\mathscr{r}}
\newcommand{\scripts}{\mathscr{s}}

\newcommand{\I}{\mathbbm{1}}
\newcommand{\R}{\mathbb{R}}
\newcommand{\Z}{\mathbb{Z}}
\newcommand{\Q}{\mathbb{Q}}
\newcommand{\C}{\mathbb{C}}
\newcommand{\F}{\mathbb{F}}
\newcommand{\T}{\mathbb{T}}
\newcommand{\Hom}{\mathrm{Hom}}

\newcommand{\Tor}{\mathrm{Tor}}
\newcommand{\Ind}{\mathrm{Ind}}
\newcommand{\ord}{\mathrm{ord}}
\newcommand{\St}{\mathrm{St}}
\newcommand{\SL}{\mathrm{SL}}
\newcommand{\GL}{\mathrm{GL}}
\newcommand{\U}{\mathrm{U}}
\newcommand{\SU}{\mathrm{SU}}
\newcommand{\DW}{\mathrm{DW}}
\newcommand{\CS}{\mathrm{CS}}
\newcommand{\WRT}{\mathrm{WRT}}
\newcommand{\llbracket}{\text{\textlbrackdbl}}
\newcommand{\rrbracket}{\text{\textrbrackdbl}}
\newcommand{\longto}{\longrightarrow}

\newtheorem{theorem}{Theorem}[section]
\newtheorem{proposition}{Proposition}[section]
\newtheorem*{conjecture*}{Conjecture}

\title{\bf Approximating $\SU(2)$ Chern-Simons theory by finite group gauge theories} 
\author[1, 2]{Thomas Nicosanti}
\author[3]{Pavel Putrov}
\author[1, 2, 3]{Johann Quenta-Raygada}
\affil[1]{SISSA, Via Bonomea 265, Trieste 34136, Italy}
\affil[2]{INFN -- Sezione di Trieste, Via Valerio 2, Trieste 34127, Italy}
\affil[3]{ICTP, Strada Costiera 11, Trieste 34151, Italy}
\date{}

\begin{document}

\maketitle

\begin{abstract}
    Motivated by some previously known facts from mathematical and physics literature, we explore certain relations between 3-dimensional topological gauge theories with continuous and finite gauge groups, commonly known as Chern-Simons (CS) and Dijkgraaf-Witten (DW) theories, respectively. Specifically, we consider the continuous and finite gauge groups to be the same algebraic group over the complex numbers and a finite field, respectively. In this paper, we focus on the $\SU(2)$ example and consider the relationship on the level of the corresponding partition functions on closed 3-manifolds. Mathematically, these are Witten-Reshetikhin-Turaev and DW invariants. We find that the asymptotics of the DW theory when the number of elements of the finite field is large recovers the leading asymptotics of the CS theory at large level when the 3-manifold contains no hyperbolic components. As a byproduct, we develop efficient techniques to count the number of points over finite fields $\mathbb{F}_q$ of $\SL(2)$ representation varieties of fundamental groups of 3-manifolds, possibly with weights pulled back from a chosen class in $H^3(\SL(2,\mathbb{F}_q),\U(1))$. 

\end{abstract}

\tableofcontents


\section{Introduction} \label{sec:intro}

Topological gauge theories, both with continuous and finite gauge groups, play an important role in condensed matter and high-energy physics. In 3 dimensions, the case of continuous gauge groups is governed by Chern-Simons theory \cite{witten1989quantum}. The partition function on a 3-manifold is formally defined via a path integral over the space of $G$-connections on the 3-manifold, for $G$ a Lie group (ordinarily assumed to be semisimple and compact). The mathematical definition of the partition function was provided by Reshetikhin and Turaev \cite{ReshetTuraev91} through the Dehn surgery representation of the 3-manifold and using the representation theory of the quantum group $\bar{U}_\xi(\mathfrak{g})$ for $\xi$ being a root of unity and $\mathfrak{g}$ the Lie algebra of $G$. The result gives a powerful topological invariant of 3-manifolds, commonly known as Witten-Reshetikhin-Turaev (WRT) invariant. The invariants of 3-manifolds and their link counterparts constructed from quantum groups are also often referred to as \textit{quantum invariants}. 

In \cite{DijkWitten90}, Dijkgraaf and Witten proposed a toy model for Chern-Simons theory in which the gauge group $G$ is a finite group. In this setting, the path integral reduces to a finite sum over flat $G$-connections and the corresponding invariants can be computed from a triangulation of the manifold. These invariants further depend on a choice of 3-cocycle $\omega \in H^3(BG, \U(1))$, also known as a twist, which plays the role of the topological action of the theory. In contrast, Chern-Simons theory for a compact simply connected simple group $G$ (such as $\SU(2)$) depends on an integer parameter $k$, known as the level. The corresponding quantum parameter in the WRT construction is\footnote{We use $k$ for the renormalized level, which differs by 2 from the classical one.} $\xi=e^{\frac{2\pi i}{k}}$. More precisely, this integer is to be understood as an element of $H^4(BG, \Z) \simeq \Z$, whose generator is the second Chern class used to define the action of the theory. Since $H^4(BG,\Z) \simeq H^3(BG, \U(1))$ for a finite group $G$, a choice of level is then analogous to a choice of twist in the discrete scenario.

In this paper, we will be interested in comparing the topological invariants (physically realized by partition functions) obtained from the continuous and discrete cases, where an appropriate choice of gauge groups (and topological actions) makes such a comparison meaningful. More specifically, we will consider the algebraic group $\SU(2)$ on the continuous side, and the same algebraic group, but over a finite field, on the discrete side. We recall that finite fields $\mathbb{F}_q$ are completely classified by their number of elements, $q$, which must be a power of a prime: $q=p^r$. The field has a Frobenius automorphism $x\mapsto x^p$ which has order $r$. Note that the definition of the ordinary $\SU(2)\equiv \SU(2,\C)$ as a subgroup of $2\times2$ invertible matrices $\GL(2,\C)$, apart from the usual algebraic operations, involves complex conjugation -- an order 2 automorphism of $\C$. Because of this, its finite field analogue $\SU(2,\mathbb{F}_{q^2})$ can be defined only over the field with $q^2$ elements, where the analogue of conjugation is played by the order 2 automorphism $x\mapsto x^q$ (in a sense, the field extension $\F_{q^2}/\F_q$ is analogous to the extension $\C/\R$). However it is known that $\SU(2,\F_{q^2})\cong \SL(2,\mathbb{F}_q)$. Similarly, one can consider $\SL(2,\C)$ as a complexification of $\SU(2)$, with $\SU(2)$ being the maximal compact subgroup. The gauge group $\SL(2,\C)$ plays a crucial role in the \textit{analytic continuation} of the $\SU(2)$ Chern-Simons theory \cite{Witten:2010cx,Kontsevich}.

A priori, it is not obvious why a relation between these two kinds of invariants should exist. One motivation comes from the celebrated Weil conjectures, which in particular imply that there is a close relationship between the algebraic varieties over complex numbers and finite fields. Namely, some topological information about the complex variety can be recovered from the count of points of the same variety over finite fields. Algebraic varieties appear naturally in the context of topological gauge theories as representation/character varieties of fundamental groups, when the group is algebraic. In particular, for the group over a finite field, the count of points of the representation variety, up to a normalization factor, equals the partition function of the corresponding untwisted DW theory. On the other hand, for the group over complex numbers, the variety is the space of critical points -- flat connections -- of the Chern-Simons functional, which plays a crucial role in the semi-classical expansion of the path-integral. We note that the $\SL(2,\F_q)$ character varieties of 3-manifolds have been studied in numerous contexts \cite{Borel1981, Harada2011, harada2014hasseweilzetafunctionssl2character, harada2019hasseweilzetafunctionsrm, LiXu2004}, but to the best of our knowledge, our work is the first instance in which a direct parallel with the $\SU(2)$ Chern-Simons theory is drawn. The count of points of $\SL(2,\F_q)$ character varieties of 2-manifolds (Riemann surfaces) was considered in particular in \cite{hausel2008mixed, Hausel2011}.

Additional motivation comes from some known examples of correspondences between finite group and continuous gauge theories. In the simplest case where $G = \U(1)\equiv \U(1,\C)\subset \GL(1,\C)$, the Chern-Simons partition function localizes to a sum over flat connections and can be computed explicitly to be (see e.g. \cite{deloup2001abelian,Guo:2018vij})
\begin{equation}
    Z_{\CS}^{\U(1),k}(M) = k^{\frac{b_1-1}{2}} \sum_{a \in \Tor H_1(M, \Z)} e^{2\pi i k \ell k(a,a)}.
    \label{intro-U1-CS}
\end{equation}
Here $b_1$ is the first Betti number of $M$ and $\ell k$ denotes the linking pairing between (torsion) 1-cycles of $M$. On the discrete side, consider $\U(1,\F_{q^2})$, which is a cyclic group $\Z/N\Z$ with $N=q+1$. Alternatively, one can consider $\GL(1,\F_q)\cong \F^\times_q\cong \Z/N\Z$ for $N=q-1$. In general, for $G=\Z/N\Z$ we have $H^3(BG, \U(1)) \simeq \Z/N\Z$, so let $k$ denote an integer corresponding to a particular choice of twist. The partition function of the Dijkgraaf-Witten theory reads (see e.g. \cite{Guo:2018vij})
\begin{equation}
    Z_{\DW}^{\Z/N\Z,k}(M) = N^{b_1-1} \sum_{\substack{a \in \Tor H_1(M, \Z) \\ o(a) \mid N}} e^{2\pi i k \ell k(a,a)},
    \label{intro-ZN-DW}
\end{equation}
where by $o(a)$ we denote the order of the torsion cycle $a$.  One explicitly observes that the discrete gauge theory encodes the same topological information as the continuous theory does, namely the first Betti number and the linking pairing evaluated on equal cycles. Moreover, when, for a given 3-manifold, $N$ is ``large enough'' in the sense that the direct limit of $\Z/N\Z$ groups gives\footnote{E.g. by considering the sequence with $N=n!$ , $n\rightarrow \infty$.} $\Q/\Z$ (cf. \cite{Putrov22}), the condition $o(a) \mid N$ is automatically satisfied. Then one can identify the expressions in the right-hand sides of (\ref{intro-U1-CS}) and (\ref{intro-ZN-DW}) by formally setting $N=k^\frac{1}{2}$.

The above results prompt us to conjecture the existence of a similar relation in the non-abelian case. In this context, the path integral does not always localize to a sum over flat connections, and so we will only consider the large $k$ limit in which such an approximation is valid. The \textit{asymptotic expansion conjecture} \cite{witten1989quantum,FreedGompf91,andersen2004asymptotic,andersen2025proofwittensasymptoticexpansion} states that the $\SU(2)_k$ Chern-Simons partition function behaves as
\begin{equation}
    Z_{\CS}^{\SU(2)_k}(M) \sim \sum_{\gamma \in \pi_0(\mathcal{M}(M, \SU(2)))} e^{2\pi i k \CS[\gamma]} \sum_{n \geq 0} \frac{a_n^{\gamma}}{k^{n + \delta_\gamma}},   
\end{equation}
as we take the limit $k \to \infty$ through integer numbers. Here $\mathcal{M}(M, \SU(2))$ denotes the moduli space of flat $\SU(2)$ connections over $M$. We can also analytically continue CS theory, i.e.\ consider $k$ to be a complex-valued level, and take the limit $k \to \infty$ in the complex plane. It has been conjectured that the corresponding asymptotic expansion must take into account not only flat $\SU(2)$ connections, but more generally flat $\SL(2,\C)$ connections \cite{Witten:2010cx,Kontsevich,gukov2016resurgence,GukovPutrov24,wheeler2025quantum} --- we will review this proposal in more detail later on (see Section \ref{sec:DWsl2q}). Alternatively, one can consider the asymptotic expansion of the WRT invariant for a general root of unity $\xi$, which formally corresponds to taking the CS level $k\in\Q$ \cite{Chen:2015wfa,wheeler2025quantum,PS2025}. Within the context of our work, these general versions of the asymptotic expansion conjecture turn out to be more relevant for a comparison with the discrete case. Note that in the abelian case, the sequence of cyclic groups $\Z/N\Z$ with $N\rightarrow\infty$ can be considered as densely filling $U(1)$. In the case of $SU(2)$, however, no such sequence of finite subgroups exists, which follows from the known ADE classification.

Our conjecture is then formulated as follows. We claim that for an appropriate choice of twist in $H^3(BG, \U(1))$, both the leading powers of $k$ in the asymptotic expansion and the twelvefold values of the Chern-Simons invariants can be recovered from the $\SL(2,\F_q)$ Dijkgraaf-Witten partition function. We do not have a general proof for these statements, except for the particular case of spherical 3-manifolds, where we propose some ideas on which a full proof could be based. Instead, we provide considerable evidence for our conjecture in a variety of examples, most of which are \textit{geometric} 3-manifolds. Interestingly enough, it is only for hyperbolic 3-manifolds --- perhaps the most interesting type of geometric 3-manifolds --- that our conjecture seems not to hold, at least not in the strongest form.

The rest of the paper is structured as follows. In Section \ref{sec:revsl2q} we introduce the finite groups $\SL(2,\F_q)$ and briefly review some of their group-theoretical aspects. We also present a particular construction of 3-cocycles due to Nosaka \cite{Nosaka23}, which we use to define suitable topological actions. In Section \ref{sec:DWsl2q} we thoroughly study the $\SL(2,\F_q)$ Dijkgraaf-Witten theories. We determine their modular data, providing a simple but powerful way to compute the partition function in certain cases. Then, having laid out all the necessary definitions, we spell out our conjecture in full detail. In Section \ref{sec:examples} we provide substantial evidence in favor of our conjecture, by computing the Dijkgraaf-Witten invariants for a variety of 3-manifolds and comparing with the $\SL(2,\C)$ Chern-Simons theory in the asymptotic limit. We also discuss an interesting (counter)example of a hyperbolic 3-manifold. Finally, in Section \ref{sec:sphmflds} we present a full proof of a weak version of our conjecture for spherical 3-manifolds.

\section{The finite groups \texorpdfstring{$\SL(2,\F_q)$}{SL(2,Fq)}} \label{sec:revsl2q}

We start by reviewing some basic facts about finite fields. Let $p$ be a prime number and $r$ a positive integer. There exists a unique (up to isomorphism) finite field of order $q = p^r$, denoted by $\F_q$. Conversely, any finite field is isomorphic to $\F_q$ for appropriate $p$ and $r$. For $r=1$ we identify $\F_p \simeq \Z/p\Z$ with the standard addition and multiplication modulo $p$, while for $r > 1$ an explicit construction is given by the quotient $\F_p[x]/(f)$, with $f$ an irreducible polynomial of degree $r$. 

The multiplicative subgroup $\mathbb{F}_q^\times$ is isomorphic to $\Z/(q-1)\Z$; we will usually denote a generator of this group by $\theta$. If $r$ divides $s$, the field $\F_{p^r}$ embeds into $\F_{p^{s}}$. It is often convenient to choose compatible generators for their multiplicative subgroups $\expval{\theta}$ and $\expval{\vartheta}$, meaning that $\theta = \vartheta^{(p^s-1)/(p^r-1)}$ after the embedding. This can be achieved by choosing the polynomials $f$ defining $\F_{p^r}$ to be e.g.\ the Conway polynomials (see for example \cite{LuxPahlings} for a detailed account). Clearly, finite fields are never algebraically closed: the algebraic closure of $\F_{p^r}$ can be obtained as the direct limit $\overline{\F}_p = \bigcup_{k = 1}^\infty \F_{p^k}$.

In what follows, we restrict ourselves to the case where $p$ is an odd prime\footnote{When $p = 2$ the general structure of the group differs slightly, but the details are not important for our purposes.}. Consider now the special linear group of $2 \times 2$ matrices over $\F_q$, 
\begin{equation}
    \SL(2,\F_q) = \qty{\mqty(a & b \\ c & d) \ \eval \ a,b,c,d \in \F_q, \ ad-bc = 1}.
\end{equation}
It has order $|\SL(2,\F_q)| = q(q^2-1)$ and center $Z = \qty{\I, -\I}$. It will also be useful to define the subgroups $B, T$, and $U$ consisting of upper triangular, diagonal, and unipotent upper triangular matrices respectively. Their elements have the general form
\begin{equation}
    u(\lambda, x) = \mqty(\lambda & x \\ 0 & \lambda^{-1}), \qquad d(\lambda) = \mqty(\lambda & 0 \\ 0 & \lambda^{-1}), \qquad u(1, x) = \mqty(1 & x \\ 0 & 1),
\end{equation}
where $\lambda \in \F_q^\times$ and $x \in \F_q$. Note that since the field $\F_q$ is not algebraically closed, there exist matrices that can be diagonalized over the quadratic extension $\F_{q^2}$ but not over $\F_q$. Their form can be written explicitly as follows. Choose an element $z \in \F^\times_{q^2}$ satisfying $z + z^q = 0$, and denote by $\mu_{q+1}$ the set of all $\xi \in \F_{q^2}^\times$ satisfying $\xi^{q+1} = 1$. For such $\xi$, define
\begin{equation}
    d'(\xi) = \pmqty{\dfrac{\xi + \xi^q}{2} & z \dfrac{\xi-\xi^q}{2} \\ \dfrac{\xi - \xi^q}{2z} & \dfrac{\xi + \xi^q}{2}}.
\end{equation}
We will informally refer to these matrices as \textit{codiagonal matrices}. They belong to $\SL(2,\F_q)$, but have eigenvalues in the quadratic extension unless $\xi = \pm 1$. We denote by $T'$ the subgroup of all matrices $d'(\xi)$ with $\xi \in \mu_{q+1}$, which is isomorphic to $\Z/(q+1)\Z$.

\subsection*{Conjugacy classes and irreducible representations}

From the above considerations one concludes that every matrix in $\SL(2,\F_q)$ is either $\pm \I$, or is conjugate to a diagonal, codiagonal, or unipotent upper triangular matrix. A more detailed study shows that there exist $q+4$ conjugacy classes, which can be characterized as follows \cite{Bonnafe}. Define equivalence relations identifying $\lambda \sim \lambda^{\pm1}$ in $\F_q^\times$ and $\xi \sim \xi^{\pm1}$ in $\F^\times_{q^2}$, and let $z_0$ be a fixed non-square element in $\F_q$. A complete set of representatives is then given by 
\begin{equation} \label{eq:sl2qconjreps}
    \begin{cases}
        \pm \I, \\
        d(\lambda), & \lambda \in (\F^\times_q/\sim) \backslash \qty{\pm 1}, \\
        d'(\xi), & \xi \in (\mu_{q+1}/\sim) \backslash \qty{\pm 1}, \\
        \epsilon u_\tau, & \epsilon = \pm 1 \ \text{and} \ \tau = \pm 1,
    \end{cases}
\end{equation}
where we defined $u_+ = u(1,1)$ and $ u_- = u(1,z_0)$. We summarize the structure of the corresponding conjugacy classes in Table \ref{tab:sl2qconj}.
\begin{table}[!ht]
    \centering
    \begin{tabular}{Sc | Sc Sc Sc Sc}
         Representative & $\pm \I$ & $d(\lambda)$ & $d'(\xi)$ & $\epsilon u_\tau$ \\
         \hline
         Number of classes & 2 & $\dfrac{q-3}{2}$ & $\dfrac{q-1}{2}$ & 4 \\
         Cardinality & $1$ & $q(q+1)$ & $q(q-1)$ & $\dfrac{q^2-1}{2}$ \\
         Centralizer & $\SL(2,\F_q)$ & $T$ & $T'$ & $ZU$ \vphantom{$\dfrac{q^2-1}{2}$}
    \end{tabular}
    \caption{Conjugacy classes of $\SL(2,\F_q)$.}
    \label{tab:sl2qconj}
\end{table}

The (irreducible) characters of $\SL(2,\F_q)$ are also well-known in the literature and are constructed as follows \cite{Reeder14}. Let $\alpha$ be a character of $\F_q^\times \simeq \Z/(q-1)\Z$. We can lift it to a character of $B$, acting as $\alpha(u(\lambda, x)) = \alpha(\lambda)$. The \textit{principal series representations} of $\SL(2,\F_q)$ are then defined as the induced representations $R(\alpha) = \Ind_B^G\alpha$. They are irreducible unless $\alpha$ is of order 2, in which case they decompose into irreps as
\begin{equation}
    R(1) = 1 + \St, \qquad R(\alpha_0) = R_+(\alpha_0) + R_-(\alpha_0),
\end{equation}
where $\alpha_0$ denotes the non-trivial character of order 2. One further checks $R(\alpha) \simeq R(\alpha^{-1})$.  

Now let $\beta$ be a character of $\mu_{q+1} \simeq \Z/(q+1)\Z$. The  \textit{cuspidal representations} $R'(\beta)$ are given as follows. For any character $\alpha \in \widehat{\F_q^\times}$ agreeing with $\beta$ on $\pm 1$, we define the virtual character $R'(\beta) = \Ind_T^G\alpha - R(\alpha) - \Ind_{T'}^G \beta$, where we understand $\alpha$ and $\beta$ as being lifted to characters of $T$ and $T'$ in the obvious way. One can show they correspond to actual representations of $\SL(2,\F_q)$, which are moreover irreducible, whenever $\beta$ is not of order 2. Otherwise, one has (virtual) decompositions into irreps
\begin{equation}
    R'(1) = 1 - \St, \qquad R'(\beta_0) = R'_+(\beta_0) + R'_-(\beta_0),
\end{equation}
where $\beta_0$ denotes the non-trivial character of order 2. Just as for principal series representations, one checks $R(\beta) \simeq R'(\beta^{-1})$.

These two constructions exhaust all irreducible representations of $\SL(2,\F_q)$. Thus, we conclude that a complete set of characters is given as follows:
\begin{equation} \label{eq:sl2qchars}
    \begin{cases}
        1, \\
        \text{St}, \\
        R(\alpha), & \alpha \in (\widehat{\F_q^\times}/\sim) \backslash \qty{\pm 1}, \\
        R'(\beta), & \beta \in (\widehat{\mu_{q+1}}/\sim) \backslash \qty{\pm 1}, \\
        R_{\sigma}(\alpha_0), & \sigma = \pm 1, \\
        R'_{\sigma}(\beta_0), & \sigma = \pm 1,
    \end{cases}
\end{equation}
where we defined an equivalence relation identifying characters $\alpha \sim \alpha^{\pm 1}$ in $\widehat{\F_q^\times}$ and $\beta \sim \beta^{\pm 1}$ in $\widehat{\mu_{q+1}}$. Their values on conjugacy class representatives are given in Table \ref{tab:sl2qchars}.
\begin{table}[!ht]
    \centering
    \begin{tabular}{Sc | Sc Sc Sc Sc}
         & $\epsilon \I$ & $d(\lambda)$ & $d'(\xi)$ & $\epsilon u_\tau$  \\
         \hline
         $1$ & 1 & 1 & 1 & 1 \\
         $\mathrm{St}$ & $q$ & 1 & $-1$ & 0 \\
         $R(\alpha)$ & $(q+1)\alpha(\epsilon)$ & $\alpha(\lambda) + \alpha(\lambda)^{-1}$ & 0 & $\alpha(\epsilon)$ \\
         $R'(\beta)$ & $(q-1) \beta(\epsilon)$ & 0 & $-\beta(\xi) - \beta(\xi)^{-1}$ & $-\beta(\epsilon)$ \\
         $R_\sigma (\alpha_0)$ & $\frac{(q+1)\alpha_0(\epsilon)}{2}$ & $\alpha_0(\lambda)$ & 0 & $\alpha_0(\epsilon) \frac{1 + \sigma \tau \sqrt{q_0}}{2}$ \\
         $R'_\sigma (\beta_0)$ & $\frac{(q-1)\beta_0(\epsilon)}{2}$ & 0 & $-\beta_0(\xi)$ & $\beta_0(\epsilon) \frac{-1 + \sigma\tau \sqrt{q_0}}{2}$ \\
    \end{tabular}
    \caption{Character table of $\SL(2,\F_q)$. Here $q_0 = \alpha_0(-1) q$.}
    \label{tab:sl2qchars}
\end{table}

\subsection{Group cohomology and second Chern classes} \label{subsec:12chern}

In this section, we proceed with a discussion of the group cohomology of $\SL(2,\F_q)$, as well as explicit constructions of representatives for group cocycles in $H^3(\SL(2,\F_q), \U(1))$. The latter will become essential in our study of $\SL(2,\F_q)$ gauge theories in Section \ref{sec:DWsl2q}. 

We start by pointing out the fact that the abelianization of $\SL(2,\F_q)$ is trivial for $q > 3$. This implies that
\begin{equation} 
    H^1(\SL(2,\F_q), \U(1)) = 0
\end{equation}
in those cases. It turns out that also
\begin{equation}
    H^2(\SL(2,\F_q), \U(1)) = 0
\end{equation}
for $q \notin \qty{4, 9}$ \cite{Schur1904}. The third group cohomology is however non-trivial and can be determined as follows. The Bockstein map $\beta$ associated to the short exact sequence 
\begin{equation} \label{eq:bockses}
    0 \longto \Z \longto \R \longto \U(1) \longto 0
\end{equation}
induces an isomorphism between $H^3(\SL(2,\F_q), \U(1))$ and $H^4(\SL(2,\F_q), \Z)$, which is in turn isomorphic to $H_3(\SL(2,\F_q), \Z)$ by the universal coefficient theorem. The third group homology was computed in \cite{Hutch13} and equals $\Z/(q^2-1)\Z$ whenever $q \notin \qty{2, 3, 4, 5, 8, 9, 27}$. Hence we conclude that also
\begin{equation}
    H^3(\SL(2,\F_q), \U(1)) = \Z/(q^2-1)\Z
\end{equation}
for such values of $q$. Throughout this work, we will make the assumption that $q$ is not any of the special values for which the group cohomology behaves differently.

Let us now introduce a procedure to find explicit group 3-cocycles constructed from second Chern classes of representations of $\SL(2,\F_q)$, following previous work by Nosaka \cite{Nosaka23}. Let $G$ be a finite group. Given a representation $\rho: G \to \GL(n,\C)$, one defines the Chern classes $c_{i}(\rho) \in H^{2i}(G, \Z)$ of $\rho$ as the Chern classes of its associated vector bundle over $BG$. Brauer's induction theorem tells us that for any representation $\rho$, there exist finitely many subgroups $H_i \subset G$ and corresponding 1-dimensional representations $\phi_i: H \to \C^\times$ such that $\rho$ decomposes, as a virtual representation, into
\begin{equation}
    \rho = \sum_{i=1}^N n_i(\rho) \Ind_{H_i}^G(\phi_i),
\end{equation}
with $n_i(\rho) \in \Z$. Proposition 2.2\ in \cite{Nosaka23} states that the twelvefold second Chern class $12\beta^{-1} c_2(\rho) \in H^3(G,\U(1))$, where $\beta$ is the Bockstein map associated to (\ref{eq:bockses}), can be written as
\begin{equation} \label{eq:12chern}
    12\beta^{-1}c_2(\rho) = 6 \sum_{k=1}^N n_k(\rho) \qty(\qty(\sum_{j=1}^N n_j(\rho) \Tr_{H_j}^G(\phi_j) \cup \beta \circ \Tr_{H_k}^G(\phi_k))- \Tr_{H_k}^G(\phi_k \cup \beta\phi_k)).
\end{equation}
Some clarifications are in order. First, note that we are expressing $12\beta^{-1}c_2(\rho)$ as a group cocycle with $\Q/\Z$ coefficients. This is possible since the group cohomology of finite groups is pure torsion, and hence the cohomologies with $\Q/\Z$, $\U(1)$, or even $\C^\times$ coefficients are all isomorphic\footnote{We will employ additive or multiplicative notation for the coefficient group interchangeably, understanding that group cocycles are expressed with coefficients in either $\Q/\Z$ or $\U(1)$ respectively.}. A 1-dimensional representation $\phi: H \to \C^\times$ can thus be interpreted as a 1-cocycle in $H^1(H, \Q/\Z) \simeq \Hom(H, \Q/\Z)$, and applying the Bockstein map yields a 2-cocycle with coefficients in $\Z$. The cup product $\cup$ is the one associated to the product $\Z \times \Q/\Z \to \Q/\Z$ sending $(m, [a]) \mapsto [ma]$. Finally, $\Tr_{H}^G$ denotes the transfer map $H^n(H,A) \to H^n(G,A)$. A formal definition can be found in \cite{Brown}, but we will present concrete expressions in what follows.

Let us briefly recall the algebraic description of group cocycles. Cochains in $C^n(G,A)$ are represented by maps $\omega:G^{n}\to A$. The coboundary operator $\delta$ is defined as
\begin{equation}
    \begin{aligned}
        \delta \omega (g_1, \hdots, g_{n+1}) = (-1)^{n+1} & \omega(g_1, \hdots, g_n) + \omega(g_2, \hdots, g_{n+1})  \\
        & + \sum_{i=1}^{n} (-1)^i \omega(g_1, \hdots, g_ig_{i+1}, \hdots, g_{n+1})
    \end{aligned}
\end{equation}
in additive notation. Elements of $H^n(G,A)$ are then represented by cocycles $\omega$ satisfying $\delta \omega = 0$, modulo coboundaries. Moreover, it is always possible to find \textit{normalized} representatives satisfying $\omega(g_1, \hdots, g_n) = 0$ whenever any one of the $g_i$ equals $1$ \cite{EilenbergMacLane47}. 

With these definitions, it is easy to describe the cohomology operations appearing in (\ref{eq:12chern}) directly on cochains. Consider a 1-cocycle $\phi:H \to \Q/\Z$. The 2-cocycle $\beta\phi$ is represented by a function $H \times H \to \Z$, which reads
\begin{equation} \label{eq:bockexplicit}
    \beta\phi(h_1, h_2) = \overline{\phi(h_1)}+\overline{\phi(h_2)}-\overline{\phi(h_1)+\phi(h_2)} = \left\{\begin{matrix} 
        1, & \text{if} \ \ \overline{\phi(h_1)}+\overline{\phi(h_2)} \geq 1, \\
        0, & \text{if} \ \ \overline{\phi(h_1)}+\overline{\phi(h_2)} < 1,
    \end{matrix} \right.
\end{equation}
where $\overline{x} \in [0, 1)$ denotes a lift of $x$ from $\Q/\Z$ to $\Q$. Now let $R$ be a complete set of right coset representatives for $H\backslash G$, such that the identity represents the trivial coset $H$. For any $g \in G$, we denote by $\overline{g}$ the representative of the coset $Hg$. Notice then that $\overline{r} g \overline{rg}^{-1} \in H$ for any $r\in R$. For $\omega \in H^n(H,A)$, applying the transfer map yields an $n$-cocycle $\Tr_H^G(\omega) \in H^n(G,A)$ represented by a function $G^{n} \to A$ as \cite{Weiss} 
\begin{equation} \label{eq:trexplicit}
    \Tr_H^G(\omega)(g_1, \hdots, g_n) = \sum_{r \in R} \omega(\overline{r} g_1 \overline{rg_1}^{-1}, \overline{rg_1}g_2\overline{rg_1g_2}^{-1}, \hdots, \overline{rg_1\cdots g_{n-1}}g_n\overline{rg_1\cdots g_n}^{-1}),
\end{equation}
again in additive notation. Combining these results allows us to find explicit representatives for the twelvefold Chern classes (\ref{eq:12chern}) as functions $G^{3} \to \Q/\Z$.

We now specialize to $G=\SL(2,\F_q)$. For simplicity, let us assume that we start with a representation $\rho$ given as a linear combination of induced representations. Since the first and second group cohomologies of $\SL(2,\F_q)$ vanish, we can always discard the first term in (\ref{eq:12chern}), considerably simplifying the expression for the twelvefold second Chern classes. Let us now focus on representations induced from those of $T$ and $T'$, which we shall use extensively throughout this work. Let $\theta$ and $\vartheta$ be compatible generators of $\F_q^\times$ and $\F_{q^2}^\times$ respectively. Consider the induced representations
\begin{numcases}{}
    \Ind_T^G \phi, & $\phi: d(\theta^j) \longto \dfrac{j}{q-1} \text{ mod 1,}$ \label{eqn:Trep} \\
    \Ind_{T'}^G \phi', & $\phi': d'(\vartheta^{(q-1)j}) \longto \dfrac{j}{q+1} \text{ mod 1}.$ \label{eqn:T'rep}
\end{numcases}
In particular, we will be interested in the linear combination
\begin{equation}
\label{eqn:rho}
    \rho = \Ind_T^G\phi + \Ind_{T'}^G\phi',
\end{equation}
whose twelvefold second Chern class reads
\begin{equation}
\label{eqn:choiceTwisting}
    12\beta^{-1}c_2(\rho) = -6 \Tr_{T}^G (\phi \cup \beta \phi) -6 \Tr_{T'}^G (\phi' \cup \beta \phi').
\end{equation}
The motivation for said choice will become evident a posteriori in Section \ref{sec:relationCS}, where we state our main conjecture.

\subsection{The transfer map on 3-cocycles}

Let us study the properties of the transfer map for some important subgroups of $\SL(2,\F_q)$, namely the centralizers $C_G(g)$ for $g \in \SL(2,\F_q)$, which are isomorphic to $T$, $T'$, $ZU$ or $\SL(2,\F_q)$ itself, depending on $g$. For our purposes, it will suffice to evaluate the transfer map $\Tr_H^G$ restricted to the centralizers. 

Fix $\psi \in H^3(H,\Q/\Z)$. The restriction of the transfer map satisfies the following crucial identity~\cite{Brown}:
\begin{equation}
\label{eqn:generalFormulaRestrictionTransfer}
    \Tr_H^G (\psi) |_{H'} = \sum_{g\in E} \Tr_{H' \cap gHg^{-1}}^{H'} ( g \star \psi |_{H' \cap g H g^{-1}}),
\end{equation}
where $E$ is a set of representatives for the double cosets in $H' \backslash G / H$ and $\star$ is the action induced by conjugation on cocycles, i.e.
\begin{equation}
    (g \star \psi)(h_1,h_2,h_3) = \psi(g^{-1} h_1g,g^{-1} h_2g, g^{-1} h_3g).
\end{equation}

In the following, $H$ and $H'$ will be either $T$, $T'$ or $ZU$. Then, notice that $H' \cap gHg^{-1}$ is equal to the center $Z$ whenever $H \ne H'$, while if instead $H = H'$, it might also happen that $H \cap gHg^{-1} = H$. Therefore, when $H \ne H'$ we have:
\begin{equation}
\label{eqn:restrictionTransferDifferentGroups}
    \Tr_H^G (\psi) |_{H'} =
    \sum_{g\in E  } \Tr_{Z}^{H'} ( g \star \psi |_Z).
\end{equation}
The other three possibilities have to be treated separately. Suppose $H = H' = T$, then $T \cap gTg^{-1} = T$ if and only if $g \in T$ or if $g$ is of the form
\begin{equation}
    g = \begin{pmatrix}
        0 &\mu\\
        - \mu^{-1} &0
    \end{pmatrix},
\end{equation}
for some $\mu \in \F_q^\times$. Notice that the latter are exactly the matrices satisfying $gd(\lambda)g^{-1} = d(\lambda^{-1})$. We shall now determine whether or not these matrices are included in $E$ and hence contribute to~\eqref{eqn:generalFormulaRestrictionTransfer}. It turns out that it is enough to include one from each family --- for example $\I$ and $s$ are a natural choice of representatives, with
\begin{equation}
    s =
    \begin{pmatrix}
        0 &1 \\
        -1 &0
    \end{pmatrix}.
\end{equation}
 Therefore, we obtain
\begin{equation}
\label{eqn:restrictionTransferT}
    \Tr_T^G (\psi) |_{T} = \psi + \psi \circ \iota +
    \sum_{g\in E \setminus \{  \I, \ s \}} \Tr_{Z}^{T} ( g \star \psi |_Z),
\end{equation}
where $\iota$ is the inversion map on triples of elements:
\begin{equation}
    \iota(h_1,h_2,h_3) = (h_1^{-1},h_2^{-1},h_3^{-1}).
\end{equation}
Similarly, we deduce the following expressions for the last two cases: 
\begin{align}
\label{eqn:restrictionTransferT'}
    &\Tr_{T'}^G (\psi) |_{T'} = \psi + \psi \circ \iota +
    \sum_{g\in E \setminus \{  \I, \ s' \}} \Tr_{Z}^{T'} ( g \star \psi |_Z), \\
    &\Tr_{ZU}^G (\psi) |_{ZU} = \sum_{\lambda \in \F_q^\times /\sim} d(\lambda) \star \psi +
    \sum_{g\in E \setminus \{  d(\lambda) \colon \lambda \in \F_q^\times /\sim \}} \Tr_{Z}^{ZU} ( g \star \psi |_Z),
    \label{eqn:restrictionTransferU}
\end{align}
where $s'$ is any matrix realizing the inversion map on $T'$ through conjugation and $\lambda \sim \lambda'$ if and only if $\lambda' \in \{\lambda, -\lambda\}$.
Finally, notice that multiplication by $2$ kills the term $\Tr_{Z}^{H} ( g \star \psi |_Z)$ for any $H$, since $H^3(\Z/2\Z,\Q/\Z)=\Z/2\Z$. This remark is crucial since we will later specialize to $\psi = -6\phi \cup \beta\phi$, which vanishes when restricted to the center.

\section{\texorpdfstring{$\SL(2,\F_q)$}{SL(2,Fq)} Dijkgraaf-Witten theory} \label{sec:DWsl2q}

Let $G$ be a finite group. A 3-dimensional discrete gauge theory with gauge group $G$ can be constructed as follows \cite{DijkWitten90}. Fix a group cohomology class $\omega$ in $H^3(G,\U(1))$ and an oriented closed 3-manifold $M$. The Dijkgraaf-Witten theory twisted by $\omega$ is defined through its partition function as 
\begin{equation} \label{eq:ZDW}
    Z_{\DW}(M) = \frac{1}{|G|} \sum_{\gamma \in \Hom(\pi_1(M), G)} \expval{\gamma^* \omega, [M]},
\end{equation}
where we use $\gamma$ to also denote the classifying map $M \to BG$ of the corresponding flat connection, and $[M]$ denotes the fundamental class in $H_3(M, \Z)$. The above expression is manifestly a topological invariant of the 3-manifold $M$. Indeed, Dijkgraaf-Witten theory is an example of a bona fide 3-dimensional TQFT \cite{Freed93}. When the cohomology class is trivial, for an algebraic $G$, the partition function simply counts the number of points in the $G$ \textit{representation variety} $\Hom(\pi_1(M), G)$ of the manifold, up to the overall normalization factor\footnote{More naturally, the ratio ${|\Hom(\pi_1(M), G)|}/{|G|}$ can be understood as the ``number of points'' in the ${\Hom(\pi_1(M), G)}/{G}$ stack. }. We note that this count is closely related, but not exactly the same, as the count of points in the \textit{character variety} $\Hom(\pi_1(M), G)/\!\!/ G$.

We now proceed to study in the detail the Dijkgraaf-Witten theories with gauge group\footnote{The modular tensor category that describes the untwisted version was studied in \cite{chen2018representationcategoryquantumdouble}. In our context, however, we will be interested in computing partition functions on closed manifolds in the twisted version.} $G = \SL(2,\F_q)$.  In order to compute the partition function (\ref{eq:ZDW}) we must determine the space of flat connections, $\Hom(\pi_1(M), \SL(2,\F_q))$. Suppose that we are given a finite presentation of the fundamental group of $M$, consisting of $m$ generators $x_i$ and $n$ relations $R_j(x_1, \hdots, x_m)$. The set of $\SL(2,\F_q)$ flat connections is in one-to-one correspondence with the set of solutions to the system
\begin{equation}  \label{eq:flatconneq}
    R_j(X_1, \dots, X_m) = \I, \qquad \forall j \in \qty{1, \dots, n},
\end{equation}
where the $X_i$ are matrices in $\SL(2,\F_q)$. This system of equations is generally hard to solve analytically. However, for simple enough cases our knowledge of the conjugacy classes of $\SL(2,\F_q)$ allows us to determine the space of flat connections and its orbits under conjugation. We will showcase several examples in which this is possible in Section \ref{sec:examples}.

The next step is to compute the topological action $\expval{\gamma^*\omega, [M]}$. To this end, we resort to an alternative definition of Dijkgraaf-Witten theory as a \textit{lattice gauge theory}. Fix a triangulation of the 3-manifold $M$ as well as an ordering of its $N$ vertices $\qty{v_1, \hdots, v_{N}}$. A \textit{$G$-coloring} of $M$ is an assignment of elements $g_{ij} \in G$ to each link $\ell_{ij}$, oriented from vertex $v_j$ to vertex $v_i$. For consistency, we define $\ell_{ji} = \ell_{ij}^{-1}$. Such an assignment defines a flat connection over $M$ if and only if the holonomies around 2-simplices are all vanishing. That is, for any 2-simplex with vertices $v_i, v_j$ and $v_k$, the equation 
\begin{equation}
    g_{ij}g_{jk}g_{ki} = 1   
\end{equation} 
holds. Given a $G$-coloring $\varphi$, we can then compute the phase $\expval{\gamma^*\omega, [M]}$ of the corresponding flat connection $\gamma$ as follows. The ordering on the vertices of the triangulation of $M$ induces an ordering on the vertices of each tetrahedron. If $\tau$ is a tetrahedron with ordered vertices $\qty{v_{i_0}, v_{i_1}, v_{i_2}, v_{i_3}}$, we assign to it the phase factor $W(\tau; \varphi) = \omega(g_{i_0i_1},g_{i_1i_2},g_{i_3i_4})$, where we understand $\omega$ as a group cochain from $G^3 \to \U(1)$. The topological action is then a product of the single contributions from each tetrahedron in the triangulation, 
\begin{equation}
    \expval{\gamma^*\omega, [M]} = \prod_{\tau} W(\tau; \varphi)^{\epsilon_\tau},
\end{equation}
where $\epsilon_\tau = +1$ if $\tau$ has the same orientation as $M$, and $-1$ otherwise. From the lattice gauge theory perspective, the Dijkgraaf-Witten partition function reads
\begin{equation} \label{eq:ZDWlattice}
    Z_{\DW}(M) = \frac{1}{|G|^{N}} \sum_{\text{$G$-colorings } \varphi} \prod_{\tau} W(\tau; \varphi)^{\epsilon_\tau}.
\end{equation}
Naturally, this object does not depend on the particular choice of triangulation of $M$ nor does it depend on the choice of representative for $\omega$ \cite{Wakui92}. Notice also that different colorings may correspond to the same flat connection, which explains the normalization factor in the partition function.

\begin{figure}[ht!]
    \centering
    \begin{subfigure}[b]{0.4\textwidth}
        \centering
        \begin{tikzpicture}
            \begin{scope}[black, thick, decoration={
                markings, mark=at position 0.5 with {\arrow{<}}
            }]
            \draw[postaction={decorate}] (0,0) -- (2.3,-0.5);
            \draw[postaction={decorate}] (2.3,-0.5) -- (3.2,0.5);
            \draw[postaction={decorate}] (3.2,0.5) -- (1.6,2.5);
            \draw[postaction={decorate}] (0,0) -- (1.6,2.5);
            \draw[postaction={decorate}] (2.3,-0.5) -- (1.6,2.5);
            \draw[postaction={decorate}, dashed] (0,0) -- (3.2,0.5);
            \end{scope}

            \node[left] at (0,0) {$i_0$};
            \node[below] at (2.3,-0.5) {$i_1$};
            \node[right] at (3.2,0.5) {$i_2$};
            \node[above] at (1.6,2.5) {$i_3$};
        \end{tikzpicture}
        \caption{}
    \end{subfigure}
    \quad
    \begin{subfigure}[b]{0.4\textwidth}
        \centering
        \begin{tikzpicture}
            \begin{scope}[black, thick, decoration={
                markings, mark=at position 0.5 with {\arrow{<}}
            }]
            \draw[postaction={decorate}, dashed] (0,0) -- (3.2,0.5);
            \draw[postaction={decorate}] (3.2,0.5) -- (2.3,-0.5);
            \draw[postaction={decorate}] (2.3,-0.5) -- (1.6,2.5);
            \draw[postaction={decorate}] (3.2,0.5) -- (1.6,2.5);
            \draw[postaction={decorate}] (0,0) -- (2.3,-0.5);
            \draw[postaction={decorate}] (0,0) -- (1.6,2.5);
            \end{scope}

            \node[left] at (0,0) {$i_0$};
            \node[below] at (2.3,-0.5) {$i_2$};
            \node[right] at (3.2,0.5) {$i_1$};
            \node[above] at (1.6,2.5) {$i_3$};
        \end{tikzpicture}
        \caption{}
    \end{subfigure}
    \caption{A tetrahedron in a branched triangulation can be of two types: (a) positively oriented, or (b) negatively oriented.}
    \label{fig:branching}
\end{figure}
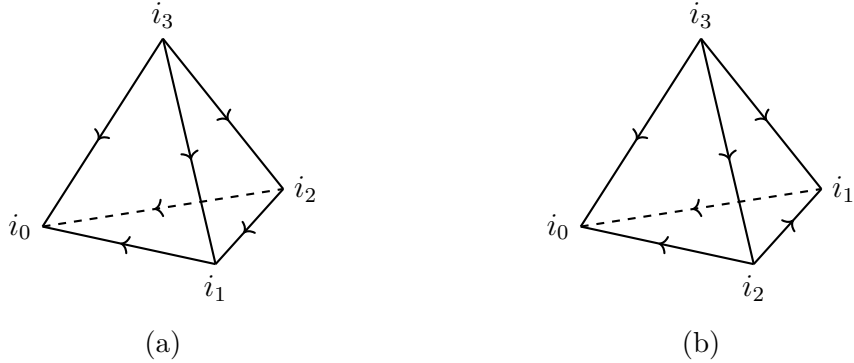

In practice, triangulations for various kinds of 3-manifolds can be obtained with the help of software packages such as Regina \cite{Regina} or SnapPy \cite{SnapPy}. However, we would like to stress that not all triangulations obtained this way are suitable for computing Dijkgraaf-Witten invariants. Indeed, ordering the vertices (and consequently the links) of a triangulation introduces an additional structure known as a \textit{branching structure}. A branching consists of a choice of orientation for the links such that no 2-simplex has a cyclic edge orientation. A tetrahedron in a branched triangulation can then be of essentially two types, shown in Figure \ref{fig:branching}.
Branched triangulations generally require a larger number of tetrahedra compared to unbranched ones, making the calculation of $Z_\DW(M)$ more demanding computationally. A way to circumvent the necessity for a branched triangulation is to consider a choice of representative for $\omega$ satisfying
\begin{equation} \label{eq:cocyclegauge}
    \omega(g, g^{-1}, h) = \omega(g, h, h^{-1}) = 1,
\end{equation}
for all $g, h \in G$ \cite{DijkWitten90}. Within this prescription, the phases $W(\tau; \phi)$ are invariant under reorderings of the vertices. Unfortunately the existence of such a representative is not always guaranteed, and depends on certain divisibility conditions which we review in Appendix \ref{app:acocycles}. For $G = \SL(2,\F_q)$, it turns out that a representative satisfying condition (\ref{eq:cocyclegauge}) can always be found for $\omega = 24\beta^{-1}c_2(\rho)$. We will therefore make use of this cocycle instead of (\ref{eq:12chern}) whenever other methods for computing $Z_\DW(M)$ are not available.

\subsection{Modular data} \label{subsec:modulardata}

In a TQFT, much information is encoded in the action of the mapping class group of the torus on its Hilbert space of quantum states $\mathcal{H}(\mathbb{T}^2)$. Recall that the mapping class group of the torus is $\SL(2,\Z)$, whose generators are usually denoted by the matrices
\begin{equation}
    S = \begin{pmatrix}
        0 & 1\\
        -1 & 0
    \end{pmatrix},
    \qquad 
    T = \begin{pmatrix}
        1 & 1\\
        0 & 1
    \end{pmatrix},
\end{equation}
satisfying $S^2 = -\I$ and $(ST)^3 = \I$. The $\SL(2,\Z)$-action on the Hilbert space $\mathcal{H}(\T^2)$ is then represented by matrices which we also denote by $S$ and $T$. The overall data of the finite-dimensional Hilbert space and the pair of $S$ and $T$ matrices acting on it is known as the \textit{modular data} of the TQFT.

Before delving into the details, we remark that $S$ and $T$ matrices will allow us to compute the DW partition function for a large class of $3$-manifolds, bypassing the problem of finding a triangulation. This is the case for plumbed manifolds, which we briefly review in Section~\ref{sec:plumbedManifolds}, and torus bundles.

After a quick review of the modular data for any finite group $G$, following \cite{coste2000finite}, we will focus on $G=\SL(2,\F_q)$. Given a normalized cocycle\footnote{In practice we often work with explicit representatives of cohomology classes, but it should be clear that every result is independent of this choice.} $\omega \in H^3(G,\U(1))$, we can always define a family
\begin{equation}
    \label{eqn:twisted-2-cocycle}
    \beta_a(g,h) = \frac{\omega(a,g,h) \omega(g,h,(gh)^{-1}a gh)}{\omega(g, g^{-1}ag,h)}
\end{equation}
of twisted 2-cocycles of $G$ for each $a \in G$, meaning they satisfy a twisted version of the cocycle condition:
\begin{equation}
    \beta_a(g,h) \beta_a(gh,k) = \beta_a(g,hk) \beta_{g^{-1}ag}(h,k).
\end{equation}
However, notice that if we restrict $g$ to be in the centralizer $C_G(a)$ of $a$, then $\beta_a$ gives rise to a 2-cocycle of $C_G(a)$.

It is important to observe that a $2$-cocycle $\beta \in H^2(G,\U(1))$ can be used to define projective $\beta$-representations, namely maps $\tilde{\rho} : G \to \text{GL}(V)$ such that $\tilde{\rho}(x) \tilde{\rho}(y) = \beta(x,y) \tilde{\rho}(xy)$. Clearly, the untwisted version recovers ordinary linear representations. Therefore, projective $\beta$-characters are analogously defined as the traces of projective $\beta$-representations.

We are now ready to describe the modular data. First, a basis for the Hilbert space $\mathcal{H}(\T^2)$ is given by pairs $(a,\tilde{\chi})$, where $a$ is a representative of a conjugacy class in $G$ and $\tilde{\chi}$ is an irreducible projective $\beta_a$-character of $C_G(a)$. We stress that this makes sense since $\beta_a \in H^2(C_G(a),\U(1))$. A special role is played by the element $0 \coloneq (\I, \chi_\text{triv})$ which corresponds to the vacuum state in $\mathcal{H}(\T^2)$.
The $S$ and $T$ matrices are expressed in terms of the data associated with the states, but the general formula is quite involved. However, we shall describe an important simplification that occurs when $\beta_a$ is a coboundary on $C_G(a)$ for any $a \in G$. 

Assuming that $\beta_a$ is cohomologically trivial, the projective characters are the same as the ordinary characters -- thus, in the basis for $\mathcal{H}(\T^2)$, we can replace the pair $(a,\tilde{\chi})$ with the pair $(a,\chi)$, with $\chi$ an irreducible character of $C_G(a)$. Furthermore, by definition there exists a 1-cochain $\varepsilon_a: C_G(a) \to \U(1)$ for which
\begin{equation}
    \beta_a(g,h) = \frac{\varepsilon_a(g) \varepsilon_a(h)}{\varepsilon_a(gh)}, \qquad \varepsilon_{x^{-1}ax} (x^{-1}gx) = \frac{\beta_a(x,x^{-1}gx)}{\beta_a(g,x)} \varepsilon_a(g),
\end{equation}
for all $g,h \in C_G(a)$ and $x \in G$. Then the modular matrices read
\begin{align}
\label{eqn:CTSmatrix}
    S_{(a,\chi),(b, \psi)} &= \frac{1}{|C_G(a)||C_G(b)|} \sum_{g \in G(a,b)} \chi(gbg^{-1})^*\psi(g^{-1}ag)^* \sigma(a \mid gbg^{-1})^*, \\
    T_{(a,\chi),(b, \psi)} &= \delta_{a,b}\delta_{\chi,\psi} \frac{\chi(a)}{\chi(\I)} \varepsilon_a(a),
    \label{eqn:CTTmatrix}
\end{align}
where we introduced $\sigma(h\mid g) = \varepsilon_h(g)\varepsilon_g(h)$ and $G(a,b) = \qty{g \in G \mid agbg^{-1} = gbg^{-1}a}$. 

A sufficient condition to have $\beta_a$ trivial is that $H^2(C_G(a),\U(1))$ vanishes for all $a \in G$.  In the case of $G = \SL(2,\F_q)$, if we suppose that $q$ is odd and different from $9$, all the relevant second cohomology groups vanish, except for $H^2(ZU,\U(1))$. Nonetheless, we shall see that all the $\beta_a$ are cohomologically trivial if we choose $\omega = 12\beta^{-1} c_2(\rho)$ as in~\eqref{eqn:choiceTwisting}.

Let us consider $\beta_a$ for $a$ a representative of some conjugacy class in $\SL(2,\F_q)$. First, suppose $a \neq \pm \I$. Then, we should consider equation~\eqref{eqn:twisted-2-cocycle} with $g,h \in C_G(a)$. In particular, notice that $\beta_a$ only depends on the restriction of the twist $\omega$ on $C_G(a)$. Then, making use of the identities~\eqref{eqn:restrictionTransferDifferentGroups},~\eqref{eqn:restrictionTransferT},~\eqref{eqn:restrictionTransferT'} and~\eqref{eqn:restrictionTransferU}, we compute:
\begin{align}
    \label{eqn:beta1}
    &\beta_{d(\lambda)} = - 6 \phi(\lambda) ( \beta\phi -  \beta\phi \circ \iota ),\\ \label{eqn:beta2}
    &\beta_{d'(\xi)} = - 6 \phi'(\xi) ( \beta\phi' - \beta\phi' \circ \iota),\\ \label{eqn:beta3}
    &\beta_{\epsilon u_\tau} = 0,
\end{align}
where $\iota(h_1,h_2) = (h_1^{-1},h_2^{-1})$ is the inversion map on pairs of elements.
From these, one can easily deduce the expressions for $\varepsilon_a$.
Indeed, if we call $\theta$ and $\vartheta$ the (compatible) generators of $\F_q^\times$ and $\F_{q^2}^\times$ respectively, we get:
\begin{align}
    \varepsilon_{d(\theta^a)} (d(\theta^b)) &= -6 \frac{\overline{a\vphantom{b}} \, \overline{b} + \overline{\vphantom{b}\! -a} \,\overline{-b}}{(q-1)^2} \mod{1},\\
    \varepsilon_{d'(\vartheta^{a(q-1)})} (d'(\vartheta^{b(q-1)})) &=-6\frac{\overline{a\vphantom{b}} \, \overline{b} + \overline{\vphantom{b}\! -a} \,\overline{-b}}{(q+1)^2} \mod{1}, \\
    \varepsilon_{\epsilon u_\tau} (\epsilon' u_{\tau'}) &= 0 \mod{1},
\end{align}
where $\overline{a}$ is a lift of $a \in \Z / n\Z$ to $\{0,1,\dots,n-1\} \subset \Z$. Notice that we have solved the problem of 
$H^2(ZU,\U(1)) \ne 0$ by choosing a twist that vanishes when restricted to $ZU$. We are left with the case $a= \pm \I$. It is easy to check that
\begin{equation}
    \beta_{\pm\I}(g,h)  = -6 \phi(\pm \I)  \Tr_T^G  \beta\phi(g,h)  - 6\phi'(\pm \I)  \Tr_{T'}^G  \beta\phi'(g,h) = 0,
\end{equation}
since $2\phi(\pm\I) = 2\phi'(\pm\I) = 0$. Therefore
\begin{equation}
    \varepsilon_{\pm\I}(g) = 0 \mod{1}. 
\end{equation}

We are now ready to write down the explicit expression for the $S$ matrix. First of all, notice that $a$ and $b$ must be commuting, otherwise the set $G(a,b)$ is empty and the corresponding $S$ matrix entries are vanishing. One possibility is that at least one of the two entries belongs to the center $Z$ of $G$. In this case we have
\begin{equation}
    S_{(\pm\I,\chi),(a, \psi)} = \frac{1}{|C_G(a)|} \chi(a)^* \psi(\pm \I)^*,
\end{equation}
where it is implicit that $\chi$ is a character of $G$, while $\psi$ is a character of $C_G(a)$. In the following, we will not specify which group each character is a character of, unless it is not clear from the context. Another non-vanishing entry of the $S$ matrix corresponds to having both $a,b \in T$ or $T'$:
\begin{equation}
    S_{(a,\chi),(b, \psi)} = \frac{1}{|C_G(a)|} \left( \chi(b)\psi(a) \sigma(a \mid b) + \chi(b^{-1})\psi(a^{-1}) \sigma(a \mid b^{-1})\right)^*.
\end{equation}
Finally, the last possibility is that both $a,b \in ZU$:
\begin{equation}
    S_{(a,\chi),(b, \psi)} = \frac{1}{|C_G(a)|} \sum_{\lambda \in \F_q^\times / \sim}\chi \!\left(d(\lambda) b d(\lambda)^{-1} \right)^* \psi \! \left(d(\lambda)^{-1} a d(\lambda)\right)^*,
\end{equation}
where $\lambda \sim \lambda'$ if and only if $\lambda' \in \{ \lambda, -\lambda\}$.

\subsection{The partition function of plumbed manifolds}
\label{sec:plumbedManifolds}

\begin{figure}[ht!]

\centering

\begin{subfigure}[t]{0.45\textwidth}
\centering
\begin{tikzpicture}[scale=0.25]

\draw[ultra thick] (5,0)  -- (9,0);
  \draw[ultra thick] (9,0) -- (13,0);
  \draw[ultra thick] (13,0) -- (16,3);
  \draw[ultra thick] (16,3) -- (19,6);
   \draw[ultra thick] (13,0) -- (16,-3);
    \draw[ultra thick] (5,0) -- (2,-3);
     \draw[ultra thick] (5,0) -- (2,3);
    \filldraw[black] (5,0) circle (8pt) node[left=5] {$[f_3,g_3]$};
    \filldraw[black] (2,3) circle (8pt) node[left=4] {$[f_2,g_2]$};
    \filldraw[black] (2,-3) circle (8pt) node[left=4] {$[f_1,g_1]$};
    \filldraw[black] (9,0) circle (8pt) node[below=3] {$[f_4,g_4]$};
    \filldraw[black] (13,0) circle (8pt) node[right=5] {$[f_5,g_5]$};
    \filldraw[black] (16,3) circle (8pt) node[right=4] {$[f_6,g_6]$};
    \filldraw[black] (19,6) circle (8pt) node[right=4] {$[f_7,g_7]$};
    \filldraw[black] (16,-3) circle (8pt) node[right=4] {$[f_8,g_8]$};
\end{tikzpicture}
\caption{An example of a plumbing graph.}
\label{fig:example-plumbing-graph}
\end{subfigure}
\hfill
\begin{subfigure}[t]{0.45\textwidth}
\centering
\begin{tikzpicture}[scale=0.25]
\begin{knot}[end tolerance=1pt]
\strand[ultra thick] (2, 0) 
  .. controls ++(90:1.5) and ++(0:-1) .. (5,2) node[pos=1,above] {$f_3$}
  .. controls ++(0:1) and ++(90:1.5) .. 
  (8,0)
  .. controls ++(90:-1.5) and ++(0:1) .. (5,-2)
  .. controls ++(0:-1) and ++(90:-1.5) .. (2, 0);
  \strand[ultra thick] (6, 0) 
  .. controls ++(90:1.5) and ++(0:-1) .. (9,2) node[pos=1,above] {$f_4$}
  .. controls ++(0:1) and ++(90:1.5) .. (12,0)
  .. controls ++(90:-1.5) and ++(0:1) .. (9,-2)
  .. controls ++(0:-1) and ++(90:-1.5) .. (6, 0);
 \strand[ultra thick] (10, 0) 
  .. controls ++(90:1.5) and ++(0:-1) .. (13,2) node[pos=1,above] {$f_5$}
  .. controls ++(0:1) and ++(90:1.5) .. (16,0)
  .. controls ++(90:-1.5) and ++(0:1) .. (13,-2)
  .. controls ++(0:-1) and ++(90:-1.5) .. (10, 0);
   \strand[ultra thick] (14, 1) 
  .. controls ++(135:1) and ++(45:-1) .. (15,4) 
  .. controls ++(45:1) and ++(135:1) .. (18,5)
  .. controls ++(135:-1) and ++(45:1) .. (17,2) node[pos=1,right] {$f_6$}
  .. controls ++(45:-1) and ++(135:-1) .. (14, 1);
     \strand[ultra thick] (14, -1) 
  .. controls ++(-135:1) and ++(-45:-1) .. (15,-4) 
  .. controls ++(-45:1) and ++(-135:1) .. (18,-5)
  .. controls ++(-135:-1) and ++(-45:1) .. (17,-2) node[pos=1,right] {$f_8$}
  .. controls ++(-45:-1) and ++(-135:-1) .. (14, -1);
    \strand[ultra thick] (17, 4) 
  .. controls ++(135:1) and ++(45:-1) .. (18,7) 
  .. controls ++(45:1) and ++(135:1) .. (21,8)
  .. controls ++(135:-1) and ++(45:1) .. (20,5) node[pos=1,right] {$f_7$}
  .. controls ++(45:-1) and ++(135:-1) .. (17, 4);
    \strand[ultra thick] (0, -5) 
  .. controls ++(135:1) and ++(45:-1) .. (1,-2) node[pos=1,left] {$f_1$}
  .. controls ++(45:1) and ++(135:1) .. (4,-1)
  .. controls ++(135:-1) and ++(45:1) .. (3,-4)
  .. controls ++(45:-1) and ++(135:-1) .. (0, -5);
      \strand[ultra thick] (0, 5) 
  .. controls ++(-135:1) and ++(-45:-1) .. (1,2) node[pos=1,left] {$f_2$}
  .. controls ++(-45:1) and ++(-135:1) .. (4,1)
  .. controls ++(-135:-1) and ++(-45:1) .. (3,4)
  .. controls ++(-45:-1) and ++(-135:-1) .. (0, 5);
   \flipcrossings{5,4,1,7,9,12,13}
\end{knot}
\end{tikzpicture}
\caption{Dehn surgery diagram corresponding to the plumbing graph on the left, if $g_i=0,\,\forall i$.}
\label{fig:example-plumbing-dehn-surgery}
\end{subfigure}
\caption{A plumbing graph and the corresponding framed link.}
\end{figure}
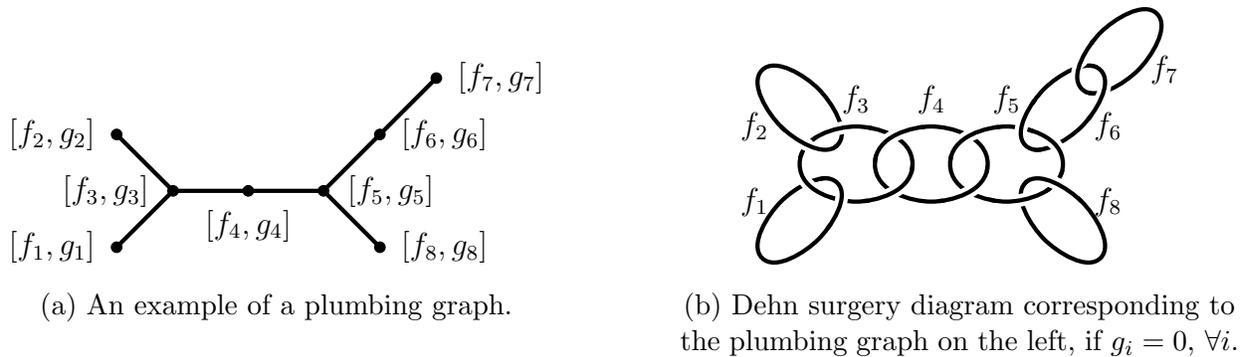

For the purposes of this work, we restrict our attention to plumbing graphs that are trees, i.e.\ connected and acyclic graphs, which are moreover closed and oriented in the sense of \cite{Neumann81}. An example of such graphs is provided in Figure~\ref{fig:example-plumbing-graph}. To each vertex $i$ of the graph $\Gamma$ we assign a non-negative integer $g_i$, the genus, and an arbitrary integer $f_i$, the weight. We may now construct a \textit{plumbed} 3-manifold from this graph as follows. To each vertex $i$ we associate a circle bundle over a genus $g_i$ Riemann surface with Euler class $f_i$. Next, to an edge between vertices $i$ and $j$ we associate the following operation. Cut out a small disk from the base space of each circle bundle --- this creates a torus boundary on the total space of each fibration. We then glue the two torus boundaries by the map $S \in \SL(2,\Z)$. We denote the resulting 3-manifold by $M_\Gamma$.

Some useful properties of the plumbed manifold $M_\Gamma$ are easily deduced from its plumbing graph $\Gamma$. For instance, the \textit{linking matrix} $B_\Gamma$ of $M_\Gamma$ is obtained as follows. The off-diagonal elements of $B$ correspond to the number of edges connecting $i$ and $j$, which are either 0 or 1 in our case. Diagonal elements are instead given by $B_{ii} = f_i$. Furthermore, one can easily deduce the first homology group of $M_\Gamma$ to be
\begin{equation}
\label{eqn:homologyPlumbed}
    H_1(M_\Gamma) = \Z^{2 \sum_i g_i} \oplus \mathrm{Coker}(B).
\end{equation}

When all genera are zero, the plumbed 3-manifold also admits an alternative representation in terms of Dehn surgery on a framed link as follows. To each vertex we assign an unknot with framing $f_i$, and two unknots form a Hopf link if the corresponding vertices are connected by an edge. Otherwise, the two unknots are unlinked (see Figure~\ref{fig:example-plumbing-dehn-surgery}).

Remarkably, the partition function of a plumbed manifold $M_\Gamma$ can be easily expressed in terms of the modular data, see~\cite{gukov2020bps}. Call $V$ and $E$ the collections of vertices and edges of $\Gamma$, then we can write
\begin{equation}
    Z_{\DW}(M_\Gamma) = \sum_{\{s(v)\}_{v \in V}} \prod_{v \in V} T^{f(v)}_{s(v),s(v)} S_{0,s(v)}^{2-2g(v)-\text{deg(v)}} \prod_{e \in E} S_{s(e_1),s(e_2)}.
    \label{eqn:plumbedDWPartitionFunction}
\end{equation}
Let us explain this expression. To each vertex $v \in V$ we associate a state $s(v) = (a, \tilde{\chi})$ in the basis of $\mathcal{H}(\T^2)$ and we sum over all possible choices. We recall that $0$ stands for the vacuum, while $\deg(v)$ is the degree of a vertex $v \in V$, i.e.\ the number of edges that end in $v$. Finally, $e_1$ and $e_2$ are the two ends of an edge $e \in E$ and $[f(v),g(v)]$ are the weight and the genus of the vertex $v$, respectively.

\subsection{Naturality of the twist} \label{sec:choiceTwist}

We will now assume that the twist $\omega$ is of the form mentioned in Section \ref{subsec:12chern}, i.e.\ $\omega = 12\beta^{-1}c_2(\rho)$, with $\rho = \Ind_T^G(\phi) + \Ind_{T'}^G(\phi')$ for $\phi$ and $\phi'$ the two $1$-dimensional representations of $T$ and $T'$ defined in~\eqref{eqn:Trep} and~\eqref{eqn:T'rep}. One crucial property of the above twist is its naturality. Indeed, consider the embedding $i_r : \SL(2,\F_q) \longto \SL(2,\F_{q^r})$, and realize $\F_q$ and $\F_{q^r}$ via Conway polynomials so that the respective multiplicative generators are compatible. If $\phi_r$ and $\phi'_r$ are the one-dimensional representations in~\eqref{eqn:Trep} and~\eqref{eqn:T'rep} corresponding to $q^r$ then $i_r^*(\phi_r) = \phi$ and $i_r^*(\phi'_r) = \phi'$. Then, we claim that $\omega$ is natural in the following sense:
\begin{equation}
    i^*_r( \omega_r ) = 12 \,i^*_r \, \beta^{-1}c_2(\Ind_{T_r}^{G_r}(\phi_r) + \Ind_{T'_r}^{G_r}(\phi_r')) =  12 \,\beta^{-1}c_2(\Ind_T^G(\phi) + \Ind_{T'}^G(\phi')) = \omega,
\end{equation}
where $G = \SL(2,\F_q)$, $G_r = \SL(2,\F_{q^r})$ and $T_r$, $T'_r$ are respectively the subgroups of diagonal and codiagonal matrices in $G_r$.
It is important to note that naturality under pullbacks does not hold in general for induced representations and transfer maps, meaning that $i_r^*$ does not simply commute with $\Ind$. However, it follows from the definition that $\Ind$ is left adjoint to the restriction functor. Consequently, the transfer map possesses certain naturality properties under restrictions in the sense of Equation~\eqref{eqn:generalFormulaRestrictionTransfer}. This is the main ingredient in the proof of the naturality of $\omega$. 

We start by considering $i^*_r \Tr_{T_r}^{G_r} ( \phi_r \cup \beta\phi_r)$. To apply~\eqref{eqn:generalFormulaRestrictionTransfer}, we need to study the double cosets $G \backslash G_r/ T_r$ and the intersection $G \cap gT_rg^{-1}$ for $g$ a representative of $G \backslash G_r/ T_r$. As one might have expected, there is a crucial difference depending on the parity of $r$. Indeed, if $r$ is even, codiagonal matrices in $G$ are similar to elements of $T_r$ in $G_r$. 
Let $r$ be odd. An element in $gT_rg^{-1}$ has order diving $q^r-1$, therefore its intersection with $\SL(2,\F_q)$ can only consist of elements of the form $h T h^{-1}$ where now $h \in \SL(2,\F_q)$ -- otherwise it equals the center. However, $G \cap g T_r g^{-1} = h T h^{-1}$ implies that either $h^{-1} g$ is in $T_r$ or it is any matrix realizing the inversion map on $T_r$ through conjugation. Either way, we can write $g = \tilde{h} \cdot t$, where $\tilde{h} \in \SL(2,\F_q)$ and $t \in T_r$, meaning that $g$ is trivial as a representative of double cosets $G \backslash G_r / T_r$. On the other hand, it is clear that for any other representative $g$ we have $G \cap g T_r g^{-1} = Z$. Then, using the fact that multiplication by $2$ kills cocycles restricted to the center, we have
\begin{equation}
    i^*_r \Tr_{T_r}^{G_r} ( 6 \phi_r \cup \beta\phi_r) = \Tr_T^G(i_r^* (6\phi_r \cup \beta \phi_r)) = \Tr_T^G( 6\phi \cup \beta \phi).
\end{equation}
The case of $r$ even is similar, except that elements of order dividing $q^r-1$ can also be of the form $h T' h^{-1}$. Call $m$ any element of $\SL(2,\F_{q^r})$ which diagonalizes codiagonal matrices in the quadratic extension, then $G \cap g T_r g^{-1} = h T' h^{-1}$ if and only if $g = \tilde{h} \cdot m \cdot  t$ for some $t \in T_r$ and $\tilde{h} \in \SL(2,\F_q)$. As above, it suffices to take $g= m$. Thus, in this case we obtain
\begin{multline}
        i^*_r \Tr_{T_r}^{G_r} ( 6 \phi_r \cup \beta\phi_r) =  \Tr_T^G( 6\phi \cup \beta \phi) + \Tr_{T'}^G( i^*_r(m \star (6\phi_r \cup \beta \phi_r))) \\
        = \Tr_T^G( 6\phi \cup \beta \phi) + \Tr_{T'}^G( 6\phi' \cup \beta \phi')
\end{multline}
Notice that there is a non-trivial mixing of the contributions of the two representations $\Ind_T^G(\phi)$ and $\Ind_{T'}^G(\phi')$. 
Finally, we have to consider the term $i^*_r \Tr_{T'_r}^{G_r} ( \phi'_r \cup \beta\phi'_r)$. It is straightforward to check that when $r$ is even, elements of $\SL(2,\F_q)$ cannot have order dividing $q^r+1$ unless they belong to the center, hence the restriction of the transfer to $\SL(2,\F_q)$ is trivial. On the other hand, when $r$ is odd $G \cap g T'_r g^{-1}$ might coincide with $h T' h^{-1}$. As above, it is enough to consider $g = \I$, so that
\begin{equation}
    i^*_r \Tr_{T'_r}^{G_r} ( 6 \phi'_r \cup \beta\phi'_r) = \Tr_{T'}^G(i_r^* (6\phi'_r \cup \beta \phi'_r)) = \Tr_{T'}^G( 6\phi' \cup \beta \phi').
\end{equation}
Putting together the contributions from the two pieces concludes the proof of $\omega = i_r^*( \omega_r)$.

One may wonder what happens when we consider a different representation $\rho$ instead of~\eqref{eqn:rho}. First, it is useful to rephrase this question in terms of the irreducible representations (\ref{eq:sl2qchars}). Notice that we may rewrite
\begin{equation}
    \Ind_T^G\phi = \St \otimes R(\phi), \qquad \qquad \Ind_{T'}^G\phi' = \St \otimes R'(\phi'),
\end{equation}
see e.g.\ Exercise 5.1 in \cite{Bonnafe}. Thus, the only contributions that we are possibly missing are those from $R_\sigma(\alpha_0)$ and $R'_\sigma(\beta_0)$. However, we expect these to contribute only for non-generic primes $p$, since they have non-trivial values only on matrices of order $p$ or $2p$, see Table~\ref{tab:sl2qchars}. 
Lastly, we remark that the choice of the particular form of $\phi$ and $\phi'$ is not canonical, since it depends on the choice of compatible generators of $\F_q^\times$ and $\F_{q^2}^\times$. This can be understood as a consequence of the freedom of acting on $\Q/\Z$ via its automorphism group.

\subsection{A conjectural relation with \texorpdfstring{$\SL(2,\C)$}{SL(2,C)} Chern-Simons theory} \label{sec:relationCS}

The main contribution of this work is a conjectural correspondence between $\SU(2)$ Chern-Simons and $\SL(2,\F_q)$ Dijkgraaf-Witten theories. By definition, the DW partition function on a closed $3$-manifold $M$ counts homomorphisms from its fundamental group to $G$, weighted by a non-trivial phase arising from the twist. In the case of $G=\SL(2,\F_q)$ Dijkgraaf-Witten theory twisted by the 3-cocycle $\omega \in H^3(G,\U(1))$ chosen above, we claim that the phase factor and the counting are closely related to the asymptotic expansion of $\SU(2)$ Chern-Simons, at least for manifolds with no hyperbolic components in their Thurston geometric decomposition.
We will provide strong evidence supporting this statement through several examples that cover a wide variety of closed oriented 3-manifolds. In Section~\ref{sec:sphmflds}, we will prove a weaker statement for spherical manifolds, which could shed some light on the relation between continuum and discrete gauge theories.

Let us be more precise. Recall that the partition function of $\SU(2)_k$ Chern-Simons can be expressed as a perturbative series as the level $k$ is sent to infinity. Indeed, it was argued in~\cite{witten1989quantum} that in such a limit the path integral localizes on the flat connections. Thus, the partition function gets a contribution from each connected component of the moduli space of flat connections $\mathcal{M}(M,\SU(2)) = \Hom(\pi_1(M),\SU(2)) / \SU(2)$. A mathematically rigorous realization of the Chern-Simons partition function is given by the WRT invariant~\cite{ReshetTuraev91}. Then, the asymptotic expansion conjecture~\cite{witten1989quantum,FreedGompf91,andersen2004asymptotic,andersen2025proofwittensasymptoticexpansion} states that the WRT invariant evaluated at a root of unity $\xi = e^{\frac{2 \pi i}{k}}$ can be expressed as
\begin{equation}
\label{eqn:wrtInteger} 
    \WRT\left(M; \xi\right) \sim \sum_{Q \in \mathcal{S}_{\SU(2)}} e^{2\pi i k Q} \, I_Q\left(\frac{1}{k}\right) ,
\end{equation}
in the limit $k \to \infty$. The sum is taken over the finite set $\mathcal{S}_{\SU(2)}$ of the different values $\CS[\gamma]$ of the Chern-Simons invariant for $\gamma$ in the moduli space of $\SU(2)$ flat connections $\mathcal{M}(M,\SU(2))$. We have also introduced $I_Q \in k^{-\Delta_Q}\C \llbracket 1/k \rrbracket$, where it is understood that the term of order $-\Delta_Q$ has non-zero coefficient. 
As the WRT invariant is well defined at any primitive root of unity $\xi = e^{\frac{2 \pi i \scripts}{\scriptr}}$, assuming that $\scripts$ is odd, the asymptotic expansion conjecture can be further refined to 
\begin{equation}
\label{eqn:wrtRational}
\WRT\left(M; \xi\right) \sim \sum_{Q \in \mathcal{S}} e^{\frac{2\pi i \scriptr }{\scripts} Q } \, I_Q^{(\scripts, \scriptr\ \text{mod} \ {\scripts})}\left(\frac{\scripts}{\scriptr}\right) ,
\end{equation}
in the limit $\scriptr \to \infty$ while keeping $\scripts$ and $\scriptr$ mod $\scripts$ fixed and $\gcd(\scriptr,\scripts)=1$. Here, the sum is taken over some finite subset $\mathcal{S}$ of $\{ z \in \C : 0 \le \Re(z) < 1\}$, while $I_Q^{(\scripts, \scriptr \ \text{mod} \ {\scripts})} \in (\scriptr/\scripts)^{-\Delta_Q}\C \llbracket \scripts/\scriptr \rrbracket$ with coefficients dependent of $\scripts$ and $\scriptr$ modulo $\scripts$. Various versions of this conjecture already appeared in the literature, see for example~\cite{Chen:2015wfa,wheeler2025quantum,PS2025}. Here, we have stated the weakest version that is needed to formulate our conjectural correspondence between Chern-Simons and Dijkgraaf-Witten theories.

We will now make some additional remarks about~\eqref{eqn:wrtRational}, and formulate a more refined version of the asymptotic expansion statement, to give some context. Similarly to how we can interpret Equation~\eqref{eqn:wrtInteger} as the asymptotics of the $\SU(2)_k$ Chern-Simons partition function, one would expect~\eqref{eqn:wrtRational} to match with the asymptotics of a Chern-Simons theory at rational level $k = \scriptr/\scripts \in \Q$. Obviously, such a Chern-Simons theory would be ill-defined as the usual action would no longer be gauge-invariant. Still, it is conjectured that one can make the partition function at rational level non-ambiguous and then interpret~\eqref{eqn:wrtRational} as its asymptotic expansion at large $\scriptr$. The set $\mathcal{S}$ would then be understood as the finite set of the different values $\CS[\gamma]$ of the Chern-Simons invariant for $\gamma$ in the moduli space of $\SL(2,\C)$ flat connections $\mathcal{M}(M,\SL(2,\C)) = \Hom(\pi_1(M),\SL(2,\C))/ \SL(2,\C)$.

Let us mention how one can make sense of $\SU(2)$ Chern-Simons theory at rational level. The idea is that Chern-Simons theory can be analytically continued, and the level $k \in \Z$ gets promoted to $k \in \C$ \cite{Witten:2010cx,Kontsevich,gukov2016resurgence}. 
The partition function can be formulated as the path integral
\begin{equation}
\label{eqn:wrtComplex}
    Z_\CS^{(\Gamma)}(M) = \int_\Gamma \mathcal{D}A \, e^{2 \pi i k \CS[A]},
\end{equation}
where $\Gamma$ is a middle-dimensional contour in the universal cover of the space of $\SL(2, \C)$ flat connections modulo gauge equivalence. When $k=\scriptr/\scripts\in \Q$ it is also sufficient to consider the $\scripts$-fold cover instead of the universal one. Notice that the gauge ambiguity of the analytically continued theory is now reflected in the ambiguity of the contour $\Gamma$. However, it is possible to choose $\Gamma$ such that it recovers the partition function of the usual $\SU(2)_k$ Chern-Simons theory, when $k$ is taken to be an integer in~\eqref{eqn:wrtComplex}. Then, if one assumes that the contour decomposes as 
\begin{equation}
    \Gamma = \sum_\alpha n_\alpha \, \Gamma_\alpha
\end{equation}
into the Lefschetz thimbles $\Gamma_\alpha$, each one corresponding to a connected component of the moduli space\footnote{More generally, Lefschetz thimbles correspond to compact middle-dimensional cycles in the connected components.},
the path integral~\eqref{eqn:wrtComplex} has a well-defined perturbative expansion. Indeed, denote by $Z_\alpha$ the path integral on the Lefschetz thimble $\Gamma_\alpha$:
\begin{equation}
    Z_\alpha =  \int_{\Gamma_\alpha} \mathcal{D}A \, e^{2 \pi i k \CS[A]}.
\end{equation}
Each path integral has an asymptotic expansion for large $k$ as follows:
\begin{equation}
    Z_\alpha = e^{2 \pi i k \CS_\alpha} \, I_\alpha\left(\frac{1}{k}\right),
\end{equation}
where $I_\alpha \in k^{-\delta_{\alpha}} \C \llbracket 1/k \rrbracket$, and $\CS_\alpha$ is the value of the CS invariant of the corresponding component of the moduli space. Notice that $\delta_\alpha$ is the leading order of $k$ in the expansion corresponding to the thimble $\alpha$. Let $\gamma_\alpha$ be any connection in the connected component of the moduli space associated with $\alpha$. It was argued in~\cite{Jeffrey92,FreedGompf91,andersen2004asymptotic,gukov2008chern} that $\delta_\alpha$ can be expressed, under certain assumptions, by the formula
\begin{equation}
\label{eqn:leadingOrderConjecture}
    \delta_\alpha = \frac{1}{2} \qty(\dim H^0(M,d+\gamma_\alpha) - \dim H^1(M,d+\gamma_\alpha)).
\end{equation}
Therefore, it is expected to be a half-integer. Moreover, if one assumes that the CS functional, considered on the space of $\SL(2,\C)$ connections modulo \textit{based} gauge transformations, behaves as a Morse-Bott function in the vicinity of the corresponding connected component $\widetilde{\mathcal{M}}_\alpha \subset \Hom(\pi_1(M),\SL(2,\C))$ in the $\SL(2,\C)$ representation variety, one can argue that~\cite{gukov2016resurgence}
\begin{equation}
    \delta_\alpha=\frac{1}{2}\left(\dim_\C SL(2,\C)-\dim_\C \widetilde{\mathcal{M}}_\alpha\right)
    \label{delta-morse-bott}
\end{equation}
Finally, this yields the following asymptotic expansion for the partition function:
\begin{equation}
\label{eqn:wrtAsympComplex}
    Z_\CS^{(\Gamma)}(M) = \sum_\alpha n_\alpha \, Z_\alpha \sim \sum_\alpha n_\alpha \, e^{2 \pi i k \CS_\alpha} \, I_\alpha\left(\frac{1}{k}\right).
\end{equation}
It was argued in~\cite{GukovPutrov24} that the integral contour $\Gamma$ might get additional contributions from Lefschetz thimbles that do not correspond to usual connected components of the moduli space of $\SL(2,\C)$ flat connections. The authors observed that these contributions arise from the non-compactness of $\SL(2,\C)$, as there might exist sequences of connections that become flat at infinity, while attaining a finite limit of the Chern-Simons functional, the so-called \textit{flat connections at infinity}.

We can now compare the asymptotic expansion~\eqref{eqn:wrtRational} of the WRT invariant at a root of unity $\xi = e^{\frac{2\pi i \scripts}{\scriptr}}$ with the asymptotics~\eqref{eqn:wrtAsympComplex} of the partition function of analytically continued $\SU(2)$ Chern-Simons at level $k = \scriptr/\scripts$. As already mentioned, one expects $\mathcal{S}$ to be the set of values of the CS functional on $\SL(2,\C)$ flat connections, and we cannot exclude that it may contain contributions from flat connections at infinity as well. Moreover, this association allows us to identify the leading orders of $I_\alpha$ and $I_Q$:
\begin{equation}
\label{eqn:leadingOrderComparison}
    \Delta_Q = \min_{\alpha : \CS_\alpha = Q} \delta_\alpha.
\end{equation}
It is well-known that the Chern-Simons functional of a manifold $M$ is valued in $\Q$ whenever $M$ does not contain any hyperbolic component in its geometric decomposition. Thus, $\mathcal{S}$ is expected to be a subset of $\Q / \Z$ in this case.

Given $\mathcal{S}$ as above, define $\mathcal{S}_\DW \subset \Q / \Z$ as the image of $\mathcal{S}$ under the map induced by multiplication by $12$ -- note that it is not injective in general. Suppose that $q$ is a power of a \textit{generic} prime, for which each element of $\mathcal{S}$ is contained either in the image of $\phi$ or in the image of $\phi'$. This condition can be equivalently stated as follows: for each $Q \in \mathcal{S}$, the order of $Q$, $o(Q)$, divides $q^2-1$. We say that $q$ is a power of a \textit{special} prime $p$, if $p$ divides $o(Q)$ for some $Q \in \mathcal{S}$. With this setting, we state the following conjecture:
\begin{conjecture*}
Given $M$ a closed oriented 3-manifold with no hyperbolic components in its geometric decomposition, the partition function of $\SL(2,\F_q)$ Dijkgraaf-Witten theory, twisted by the 3-cocycle $k\omega$, where $\omega$ is as in the beginning of this section, can be written as
\begin{equation}
\label{eqn:DWconjecture}
    Z_\DW(M)=\sum_{Q \in \mathcal{S}_\DW}  e^{2\pi i kQ} \, T_Q(q),
\end{equation}
where $T_Q$ is a function of $q$ valued in $|\SL(2,\F_q)|^{-1}\Z$ such that for large $q$ we have
\begin{equation}
    T_Q \sim q^{-2\Delta'_Q}, \qquad \Delta'_Q = \min_{Q' \in \mathcal{S} \,: \, 12Q' =Q} \Delta_{Q'}.
    \label{q-leading-power}
\end{equation}
\end{conjecture*} \noindent
Notice this is consistent with the conjectural (\ref{delta-morse-bott}) and the fact that the number of points of a variety of dimension $n$ over $\F_q$ grows as $\sim q^n$. We remark that in all the examples we have considered $|\SL(2,\F_q)| \,T_Q(q)$ is a polynomial in $q$.
Furthermore, it is implicit in~\eqref{eqn:DWconjecture} that $\mathcal{S}_\DW$ does not contain the eventual contributions from flat connections at infinity.
We stress that the condition that $o(Q)$ divides $q^2-1$ for any $Q \in \mathcal{S}$ can be relaxed --- however, it is crucial that we avoid special primes. In that case, only some elements of $\mathcal{S}_\DW$ are detected by the Dijkgraaf-Witten partition function, and the DW functional computes $Q \in \mathcal{S}_\DW$ only up to a multiple, depending on $\gcd(q^2-1,o(Q))$. We note, that as in the abelian case reviewed in Section \ref{sec:intro}, one can formally identify $q\sim k^\frac{1}{2}$ to match the leading terms in the expansions of CS and DW partition functions.

We remark that hyperbolic manifolds, which are excluded by the assumptions in the conjecture, constitute the richest and most interesting class of geometric 3-manifolds. However, there is no hope that our conjecture holds for such manifolds in its current form: the Chern-Simons functional at flat connections is in general complex-valued, and Dijkgraaf-Witten theory cannot reproduce non-rational phases. In the following section, we will consider a hyperbolic manifold as a concrete counterexample and show that there are no evident relations between the partition functions of the two theories. However, we will notice an interesting correspondence between the representation varieties over $\C$ and $\F_q$ for some specific values of $q$, and we will match the leading order $\Delta_Q$ with $\Delta'_{Q'}$.

Finally we remark on another potential direction of study of $\SL(2,\F_q)$ DW partition functions. In the untwisted case we have the following interpretation in terms of Behrend's generalization of the Grothendieck–Lefschetz trace formula:
\begin{equation}
    Z_{\DW}(M)= \# X({\mathbb{F}_{q}})=\sum_{i}(-1)^i\Tr_{H^i_c(X(\overline{\mathbb{F}_{q}}); \Q_\ell)}  \,\mathrm{Fr}^{r},
\end{equation}
where $X$ is the algebraic stack $\Hom(\pi_1(M),\SL(2))/\SL(2)$ and $\text{Fr}$ is the induced action of the Frobenius automorphism on the cohomology. When the theory is twisted by $r\omega$ it is then suggestive to ask about existence of a rank 1 sheaf $\mathcal{F}$ with $\text{Fr}|_{\mathcal{F}_\gamma}=\expval{\gamma^* \omega, [M]}$ such that for the twisted version we have
\begin{equation}
    Z_{\DW}(M)\stackrel{?}{=}\sum_{i}(-1)^i\Tr_{H^i_c(X(\overline{\mathbb{F}_{q}}); \mathcal{F})}  \,\mathrm{Fr}^{r}.
\end{equation}
In this way $H^*_c(X(\overline{\mathbb{F}_{q}}); \mathcal{F})$ would provide a categorification of the DW invariant of 3-manifolds. It would be also interesting to study the analytic properties of the corresponding zeta function, which would be a generating function for the DW partition functions for different $q=p^r$.

\section{Examples} \label{sec:examples}

In order to support our conjecture, we shall present several computations of the $\SL(2, \F_q)$ Dijkgraaf-Witten partition function on different kinds of 3-manifolds. In what follows, for simplicity, we consider the twist to be given by $\omega$ defined in Section \ref{sec:choiceTwist} itself, rather than a multiple of it.
To cover a wide variety of manifolds, we will consider at least one example for each one of the eight Thurston geometries. See Table~\ref{tab:examples} for a comprehensive list of the examples with global geometry to be discussed throughout this section.

\begin{table}[ht!]
    \centering
    \begin{tabular}{|Sc | Sc | Sc|}
        \hline
        Global geometry & Examples & Section(s) \\
        \hline
        $S^3$ & \parbox[c]{8cm}{\centering Lens spaces, Poincaré homology sphere} & \ref{sec:lensspaces}, \ref{sec:seifertHS}. See also \ref{sec:sphmflds} \\
        \hline
        $E^3$ & --- & See below \\
        \hline
        $\mathbb{H}^3$ & $-\frac{1}{2}$-surgery on the figure-eight knot & \ref{sec:hyperbolicEx}\\
        \hline
        $S^2 \times \R$ & --- & See below \\
        \hline
        $\mathbb{H}^2\times \R$ & $M(-1;(5,-2),(5,-2),(5,-1))$ & \ref{sec:h2xrExample} \\
        \hline
        $\widetilde{\SL(2,\R)}$ & \parbox[c]{8cm}{\centering Brieskorn homology spheres ($\neq \Sigma(2,3,5)$), degree-$n$ circle bundles over $\Sigma_{g > 1}$} & \ref{sec:seifertHS}, \ref{sec:higherGenus}\\
        \hline
        Nil & $T^n$ torus bundles & \ref{sec:torusBundles}\\
        \hline
        Sol & $T^nS$ torus bundles & \ref{sec:torusBundles} \\
        \hline
    \end{tabular}
    \caption{Examples treated for each of the eight Thurston geometries.}
    \label{tab:examples}
\end{table}

Notice that we did not include any example of Euclidean or $S^2 \times \R$ geometry. In both cases, there are only a few compact manifolds with that specific global geometry. 
For Euclidean geometry, five out of the six possible manifolds are torus bundles of a finite-order element of $\SL(2,\Z)$. They are left out of the analysis since two large families of torus bundles are already considered in Section~\ref{sec:torusBundles}. 
For $S^2\times \R$ geometry, the only possible manifolds are $S^2\times S^1$, $ S^2 \mathbin{ \stackrel{\sim}{\smash{\times}\rule{0pt}{0.9ex}} } S^1$ and $\R \mathbb{P}^3\#\R\mathbb{P}^3$. The first two are, respectively, the torus bundles of the identity map and the antipodal map of $S^2$, while in the last case it is enough to consider $\R\mathbb{P}^3$, which is a lens space. 
These cases are uninteresting, and nothing remarkable happens.

Recall that one way to state Thurston's geometrization conjecture is that any closed prime three-manifold has a decomposition along tori into pieces admitting one of the eight possible geometric structures. Thus, it is not enough to check our conjecture on manifolds with global geometries. 
In general, a closed oriented three-manifold whose geometric decomposition has no hyperbolic pieces is called a graph manifold. In~\cite{Neumann81} Neumann showed that graph manifolds can be obtained by plumbing. However, Neumann's definition of a plumbed manifold also allows for plumbing graphs with cycles. 
For completeness, in Section~\ref{sec:noGlobal}, we have included two general examples of plumbed manifolds with a non-trivial decomposition into Seifert fibered spaces.

\subsection{Lens spaces} \label{sec:lensspaces}

Let $M$ be the lens space $L(n,1)$, which is defined as follows. Embed the 3-sphere $S^3$ into $\C^2$, and consider the free $\Z/n\Z$ action sending 
\begin{equation*}
    (z_1, z_2) \longmapsto (e^{\frac{2\pi i}{n}} z_1, e^{\frac{2\pi i}{n}} z_2).    
\end{equation*}
The lens space $L(n,1)$ is then given as the quotient $S^3/(\Z/n\Z)$, and its fundamental group is
\begin{equation}
    \Z/n\Z = \expval{x \mid x^n = 1}.
\end{equation}
For simplicity, we will consider $n$ to be an odd prime, but our results can be easily generalized to arbitrary $n$.

\subsubsection*{Asymptotic expansion of $\SL(2,\C)$ Chern-Simons theory}

The moduli space of $\SL(2,\C)$ flat connections in this case is easily identified with the conjugacy classes of matrices of order $n$. Note that their images necessarily fall into $\SU(2)$, since upper triangular matrices have infinite order. The moduli space then consists of $\frac{n+1}{2}$ isolated points; one of them corresponds to the trivial flat connection $\gamma_0$, while the other $\frac{n-1}{2}$ correspond to the representations
\begin{equation}
    \gamma_\ell: \Z/n\Z \longto \SU(2), \quad  x \longmapsto d(\zeta_n^\ell), \qquad \ell = 1, \hdots, \frac{n-1}{2},
\end{equation}
where we defined $\zeta_n = e^{\frac{2\pi i}{n}}$. Under conjugation, their orbits are all isomorphic to $\C^\times$ (except for the trivial connection, whose orbit is a single point). From these considerations, one concludes that the powers of $k$ appearing in the asymptotic expansion are $\delta_{\gamma_0} = 3/2$ and $\delta_{\gamma_\ell} = 1/2$. Furthermore, the values of the CS functional on each flat connection read \cite{KirkKlassen90, FreedGompf91}
\begin{equation}
    \CS[\gamma_\ell] = \frac{\ell^2}{n}.
\end{equation}

\subsubsection*{$\SL(2,\F_q)$ Dijkgraaf-Witten theory}

The space of $\SL(2,\F_q)$ flat connections is now identified with the set of matrices in $\SL(2,\F_q)$ of order $n$. First, it is easy to see that $n$ should divide either one of $q-1,\ q+1$ or $q$ in order to find non-trivial flat connections. That is because matrices of type $d(\lambda)$ (respectively $d'(\xi)$, $\epsilon u_\tau$) have order dividing $q-1$ (respectively $q+1,\ p \cdot \ord(\epsilon)$). Notice that these three conditions are mutually exclusive, since $n > 2$. It is also clear that $p = n$ is a special prime, so let us focus on the other two cases.

If $n \mid q-1$, the equation $d(\lambda)^n = \I$, $\lambda^2 \neq 1$ has $\frac{n-1}{2}$ distinct solutions up to conjugation. In terms of a generator $\theta$ of $\F_q^\times$, they are
\begin{equation}
    d(\theta^{\ell \frac{q-1}{n}}), \qquad \ell = 1, \hdots, \frac{n-1}{2}.
\end{equation}
Similarly, when $n \mid q+1$, the equation $d'(\xi)^n = \I$, $\xi^2 \neq 1$ has $\frac{n-1}{2}$ distinct solutions up to conjugation. In terms of a generator $\vartheta$ of $\F_{q^2}^\times$, they read
\begin{equation} 
    d'(\vartheta^{\ell \frac{q^2-1}{n}}), \qquad \ell = 1, \hdots, \frac{n-1}{2}.
\end{equation}
However, the orbit size differs between each case, being $q(q+1)$ for the connections of $d$ type and $q(q-1)$ for the ones of $d'$ type (see Table \ref{tab:sl2qconj}).

The partition function (\ref{eq:ZDW}) is easy to find in terms of the $S$ and $T$ matrices defined in Section \ref{subsec:modulardata}. The lens space $L(n,1)$ has a simple surgery description as the $-n$-surgery on the unknot. Hence,
\begin{equation} \label{eq:lensZDW}
    Z_\DW(L(n,1)) = \sum_{s} S_{0,s}^2 \,T_{s,s}^{-n} = \frac{1}{q(q^2-1)} \qty(1 + q(q \mp 1)\sum_{\ell=1}^{\frac{n-1}{2}} e^{2\pi i \frac{12 \ell^2}{n}}), \qquad n \mid q \pm 1.
\end{equation}
Comparing with the complex case, we observe that the phase factors in the finite field partition function exactly reproduce the twelvefold Chern-Simons invariants $12 \CS[\gamma_\ell]$. Morever, the polynomials in $q$ multiplying these phases behave at leading order as $T_{\gamma_0}(q) \sim q^{-3}$ and $T_{\gamma_\ell}(q) \sim q$, in agreement with the conjectured relation $T_\gamma(q) \sim q^{-2\Delta_\gamma}$.

\subsection{Torus bundles}
\label{sec:torusBundles}

Consider a surface $\Sigma$ and an orientation-preserving diffeomorphism $\beta: \Sigma \to \Sigma$. The \textit{torus bundle} $\Sigma_\beta$ is constructed by gluing the two ends of the manifold $\Sigma \times [0,1]$ via $\beta$, as
\begin{equation}
    \Sigma_\beta = \frac{\Sigma \times [0,1]}{(x,0) \sim (\beta(x), 1)}.
\end{equation}
We will consider $\Sigma$ to be the torus $\T^2$, in which case orientation-preserving diffeomorphisms are determined up to isotopy by a matrix $U \in \SL(2,\Z)$. Note also that two torus bundles $\Sigma_U$ and $\Sigma_{U'}$ are diffeomorphic if and only if $U$ is conjugate to ${U'}^{\pm 1}$ in $\GL(2,\Z)$. If we let $U = \mqty(a & b \\ c & d)$, the fundamental group of $\Sigma_U$ can be easily written as
\begin{equation}
    \pi_1(\Sigma_U) = \expval{x, y, z \mid [x,y]=1,\ zxz^{-1} = x^ay^b,\ zyz^{-1} = x^cy^d}.
\end{equation}

The $\SU(2)$ Chern-Simons theory on torus bundles has been studied extensively in the literature \cite{KirkKlassen90, Jeffrey92}. Non-trivial values of the CS functional are well-known (at least for flat connections of the type we will study in this section), as is the asymptotic behavior of the CS partition function. The latter can actually be computed exactly by using the $S$ and $T$ matrices of the representation category of $\bar{U}_\xi(\mathfrak{sl}_2)$ for a primitive $\scriptr$-th root of unity $\xi=e^{\frac{2\pi i \scripts}{\scriptr}}$:
\begin{equation} \label{eq:su2modularmat}
    \begin{aligned}
        & S_{jl} = \frac{1}{2i}\sqrt{\frac{2}{\scriptr}} \qty(\xi^{\frac{jl}{2}}-\xi^{-\frac{jl}{2}}), \\
        & T_{jl} = \delta_{jl} e^{-\frac{\pi i}{4}} \xi^{\frac{j^2}{4}}.
    \end{aligned}
\end{equation}
For $\scripts=1$ this realizes the $\SU(2)$ Chern-Simons theory at (renormalized) integral level $k=\scriptr$.
The indices $j, l$ above run from $1$ to $\scriptr-1$, the dimension of the torus Hilbert space. Then, to compute the partition function of a torus bundle with gluing map $U$, we simply compute its trace.

For the purposes of our conjecture, we would like to compare with the asymptotics of the WRT invariant at rational levels, which as argued in Section \ref{sec:DWsl2q} includes $\SL(2,\C)$ connections. However, in contrast to the integral level case, in which only $SU(2)$ flat connections appear, we generally do not have complete information on the single contributions to the asymptotic expansion coming from each $\SL(2,\C)$ flat connection. 
This is of course not necessary in the context of our conjecture, but it could be instructive to analyze how a stronger statement may apply for these particular 3-manifolds.

We will focus on essentially two types of torus bundles, constructed via diffeomorphisms of the form $U = T^n$ and $U = T^nS$. The reason for these choices is that of studying 3-manifolds with different Thurston geometries, as mentioned at the start of this section. Torus bundles $\Sigma_U$ have Nil geometry if $U$ is a Dehn twist (such as $T^n$), while they have Sol geometry if $U$ is an Anosov map (i.e.\ if $U$ has two real distinct eigenvalues, such as $T^nS$ with $|n| > 2$). Note that the $\Sigma_{T^n}$ torus bundles can also be understood as higher-genus plumbings (see Section \ref{sec:higherGenus}), more concretely as plumbed manifolds constructed from graphs with one single vertex of genus $1$ and weight $n$. 

\subsubsection{Nil torus bundles}

We first focus on the nilmanifolds $\Sigma_{T^n}$. Let us consider $n$ to be an odd prime, although the case of general $n$ is straightforward to treat. The fundamental group simplifies to
\begin{equation}
    \pi_1(\Sigma_{T^n}) = \expval{x,y,z \mid [x,y] = 1, [z,y]=1, zxz^{-1}=xy^n}.
\end{equation}
In what follows, we denote by $X, Y$ and $Z$ the images of the generators in $\SL(2,\C)$ or $\SL(2,\F_q)$, depending on the context.

\subsubsection*{Asymptotic expansion of $\SL(2,\C)$ Chern-Simons theory}

The moduli space of flat $\SL(2,\C)$ connections can be determined analytically quite easily. The possible solutions (up to conjugacy) to the fundamental group relations are shown in Table \ref{tab:torusTnsl2c}, ordered by first fixing the conjugacy class of $X$. 
\begin{table}[ht!]
    \centering
    \begin{tabular}{Sc | Sl}
         Conjugacy class of $X$ & Possible solutions \\
         \hline
        $X = \pm \I$ & $Y = \I, \quad Z \in \SL(2,\C)$ \\
         & $Y=d(\zeta_n^\ell), \quad Z = d(*)$ \\
        \hline
        $X = d(\lambda)$ & $Y = \I, \quad Z = d(*)$ \\
         & If $\lambda = i$:\, $Y= -\I, \quad Z = \mqty(0 & * \\ -*^{-1} & 0)$ \\
         & $Y = d(\zeta_n^\ell), \quad Z = d(*)$ \\
        \hline
        $X = u(\epsilon, 1)$ & $Y = \I, \quad Z = u(\pm 1, *)$
    \end{tabular}
    \caption{Moduli space of flat $\SL(2,\C)$ connections on $\Sigma_{T^n}$.}
    \label{tab:torusTnsl2c}
\end{table}
Note that there exist families of flat connections with upper triangular images, meaning they are not $\SU(2)$ connections. Their CS invariants are however trivial. Indeed, this is true for all the flat connections with $Y = \I$: either they are abelian, or they can be continuously connected to abelian connections. Remember that for such connections the CS invariants are given by the linking form of $\Sigma_{T^n}$, which only depends on the torsion part of the homology. From the presentation of the fundamental group we find $\Tor H_1 = \Z/n\Z = \expval{y \mid y^n = 1}$, implying that when $Y = \I$ the CS invariant vanishes.  

Instead, the non-trivial values of the CS functional read
\begin{equation} \label{eq:TnCSvalues}
    \CS[\gamma] \in \qty{-\frac{\ell^2}{n} \ \eval \ \ell = 1, \hdots, \frac{n-1}{2}} \cup \qty{\frac{n}{4}}.
\end{equation}
The first $\frac{n-1}{2}$ values correspond to flat connections where $Y=d(\zeta_n^\ell)$, while the last one corresponds to the unique (up to conjugacy) flat connection with $Y = -\I$. Note already that the latter will be invisible to the discrete gauge theory, since $12\CS[\gamma]$ vanishes in that case. 

On the other hand, we do not have a way to obtain $\delta_\gamma$ for each individual flat connection. Instead, we will compute the partition function exactly at rational levels $\scriptr/\scripts$ using the $S$ and $T$ matrices in Eq.\ (\ref{eq:su2modularmat}), with $\xi = e^{\frac{2 \pi i \scripts}{\scriptr}}$, and study its asymptotic expansion as $\scriptr \to \infty$. This then determines the exponents $\Delta_Q$ associated to each CS value (see Eq.\ (\ref{eqn:wrtRational})). Assuming that $\scripts$ and $\scriptr$ are coprime and odd, and that $\scriptr$ is fixed mod $\scripts$, the asymptotic behavior of the WRT invariant reads
\begin{equation}
    \WRT(\Sigma_{T^n}; \xi) \sim e^{-\frac{\pi i n}{4}}\sqrt{\frac{i \scriptr}{2 n \scripts}} \sum_{\ell = 0}^{n-1} e^{-2\pi i \frac{\scriptr}{\scripts} \frac{\ell^2}{n}} \qty(\sum_{\lambda = 0}^{\scripts-1} e^{-2\pi i \frac{\scriptr}{\scripts} \qty(\lambda^2 n + 2\lambda \ell)}) - \frac{1}{2} e^{2\pi i\frac{\scriptr}{\scripts} \frac{n}{4}} \qty(e^{2 \pi i \frac{\scriptr}{\scripts} \frac{\scripts^2-1}{4}}).
\end{equation}
This result is in agreement\footnote{Note that the function accompanying the phase factor $e^{2\pi i \frac{\scriptr}{\scripts} \frac{n}{4}}$ is well defined for $\scriptr$ mod $\scripts$, since $\scripts$ is odd.} with the asymptotic expansion conjecture of Eq.\ (\ref{eqn:wrtRational}). For the CS value of $-\frac{\ell^2}{n}$ we find $\Delta_Q = -1/2$, as well as for the trivial CS value. For $\CS = \frac{n}{4}$, we find $\Delta_Q = 0$ instead.

\subsubsection*{$\SL(2,\F_q)$ Dijkgraaf-Witten theory}

The possible solutions to the fundamental group relations in the finite case are shown in Table \ref{tab:torusTnsl2q}, again ordered by first fixing the conjugacy class of $X$. Note we have excluded the special prime $p = n$ from our analysis.
\begin{table}[ht!]
    \centering
    \begin{tabular}{Sc | Sl}
         Conjugacy class of $X$ & Possible solutions \\
         \hline
        $X = \pm \I$ & $Y = \I, \quad Z \in \SL(2,\F_q)$ \\
         & $Y=d(\theta^{\ell \frac{q-1}{n}}), \quad Z = d(*)$ \\
         & $Y=d'(\vartheta^{\ell \frac{q^2-1}{n}}), \quad Z = d'(*)$ \\
        \hline
        $X = d(\lambda)$ & $Y = \I, \quad Z = d(*)$ \\
         & If $4 \mid q-1,\ \lambda = \theta^{\frac{q-1}{4}}$:\, $Y= -\I, \quad Z = \mqty(0 & * \\ -*^{-1} & 0)$ \\
         & If $n \mid q-1$:\, $Y = d(\theta^{\ell \frac{q-1}{n}}), \quad Z = d(*)$ \\
        \hline
        $X = d'(\xi)$ & $Y = \I$, and $Z = d'(*)$ \\
         & If $4 \mid q+1,\, \xi =  \vartheta^{\frac{q^2-1}{4}}$:\, $Y = -\I, \quad Z = A \mqty(0 & \eta \\ -\eta^{-1} & 0) A^{-1}$ \\
         & If $n \mid q+1$:\, $Y = d'(\vartheta^{\ell \frac{q^2-1}{n}}), \quad Z = d'(*)$ \\
        \hline
        $X = \epsilon u_\tau$ & $Y = \I$, and $Z = u(\pm 1, *)$
    \end{tabular}
    \caption{Moduli space of $\SL(2,\F_q)$ connections on $\Sigma_{T^n}$. Here $A$ is a matrix in $\SL(2,\F_{q^2})$ diagonalizing $d'(\xi)$, and $\eta \in \F^\times_{q^2}$ such that $Z \in \SL(2,\F_q)$.}
    \label{tab:torusTnsl2q}
\end{table}
By direct comparison, all flat connections have their analog in the complex case if we assume $n \mid q^2-1$. Observe also that $4$ always divides either one of $q-1$ or $q+1$, so there is always a single flat connection with $Y= -\I$ for any value of $q$. Moreover, we highlight the existence of connections with image in $ZU$, corresponding to the upper triangular $\SL(2,\C)$ connections.

The partition function is straightforward to compute, being equal to the trace of the gluing matrix $T^n$:
\begin{equation} \label{eq:ZDWtorusTn}
    Z_\DW(\Sigma_{T^n}) = q + 5 + (q \pm 1) \sum_{m = 1}^{\frac{k-1}{2}} e^{-2\pi i \frac{12m^2}{n}}, \qquad n \mid q \pm 1.
\end{equation}
We find that the finite field partition function reproduces well the (twelvefold) $\CS$ values in (\ref{eq:TnCSvalues}), in agreement with our conjecture. The leading powers of the polynomials $T_Q(q)$ above are also in agreement with the values of $-2\Delta_Q$, as expected.

\subsubsection{Sol torus bundles}

We now focus on the solvmanifolds $\Sigma_{T^nS}$. In this case, we will only require $n$ to be an odd number for simplicity. The fundamental group reads
\begin{equation}
    \pi_1(\Sigma_{T^nS}) = \expval{x,y,z \mid [x,y] = 1, zxz^{-1} = x^n y^{-1}, zyz^{-1} = x}.
\end{equation}
Again, we denote by $X, Y$ and $Z$ the images of the generators in $\SL(2,\C)$ or $\SL(2,\F_q)$ accordingly.

\subsubsection*{Asymptotic expansion of $\SL(2,\C)$ Chern-Simons theory}

Again in this case the moduli space can be determined analytically. The possible solutions (up to conjugacy) to the fundamental group relations are shown in Table \ref{tab:torusTnSsl2c}, ordered by first fixing the conjugacy class of $X$. 
\begin{table}[ht!]
    \centering
    \begin{tabular}{Sc | Sl}
         Conjugacy class of $X$ & Possible solutions \\
         \hline
        $X = \pm \I$ & If $X=\I$:\, $Y = \I, \quad Z \in \SL(2,\C)$  \\
        \hline
        $X = d(\lambda)$ & If $\lambda = \zeta^\ell_{n-2}$:\, $Y=X,\quad Z = d(*)$\\
         & If $\lambda = \zeta^\ell_{n+2}$: $Y=X^{-1}, \quad Z = \mqty(0 & * \\ -*^{-1} & 0)$ \\
        \hline
        $X = u(\epsilon, 1)$ & If $\epsilon = 1$:\, $Y = u(1, \alpha^{-2}), \quad Z = u(\alpha, *)$,\, with $\alpha^2 + \alpha^{-2} = n$
    \end{tabular}
    \caption{Moduli space of flat $\SL(2,\C)$ connections on $\Sigma_{T^nS}$.}
    \label{tab:torusTnSsl2c}
\end{table}
Just as in the Nil case, we can again argue that the flat connections with upper triangular images, as well as those with $X = \I$, have trivial CS invariant. Following the same line of thought, these connections are either abelian or continuously connected to abelian connections, hence their CS invariants are given by the linking form of $\Sigma_{T^nS}$. From the presentation one finds $\Tor H_1 = \Z/(n-2)\Z = \expval{x \mid x^{n-2} = 1}$, which allows us to conclude that their CS values vanish. The non-trivial values of the CS functional on $\SU(2)$ connections are instead given as follows:
\begin{equation}
    \CS[\gamma] \in \qty{ -\frac{\ell^2}{n - 2} \ \eval \ \ell = 1, \hdots, \frac{n-3}{2}} \cup \qty{ -\frac{\ell^2}{n + 2} \ \eval \ \ell = 1, \hdots, \frac{n+1}{2}}, 
\end{equation}
the first group corresponding to flat connections where $X = d(\zeta_{k-2}^\ell)$, and the second group to those where $X = d(\zeta_{k+2}^\ell)$. 

As for the leading order exponents, it is known that for all $\SU(2)$ connections one has $\delta_\gamma = 0$ (see Prop.\ 5.6(a) in \cite{Jeffrey92}), but unfortunately we cannot determine them for $\SL(2,\C)$ connections. We then proceed as in the Nil case by computing the large $\scriptr$ behavior of the WRT invariant for rational levels. With the same assumptions --- namely that $\scriptr$ and $\scripts$ are coprime and odd, and that $\scriptr$ is fixed mod $\scripts$ --- we find
\begin{equation}
    \WRT(\Sigma_{T^nS}; \xi) \sim e^{-\frac{\pi i n}{4}} \sum_{\pm} \pm \frac{1}{2\sqrt{i(n \pm 2)}} \sum_{\ell = 0}^{(n\pm 2) - 1} e^{-2\pi i\frac{\scriptr}{\scripts} \frac{\ell^2}{n\pm 2}} \qty(\frac{1}{\sqrt{\scripts}}\sum_{\lambda=0}^{\scripts-1} e^{-2\pi i \frac{\scriptr}{\scripts} \qty(\lambda^2 (n+2) + 2\lambda \ell)}),
\end{equation}
which again is in agreement with the asymptotic expansion conjecture of Eq.\ (\ref{eqn:wrtRational}). From here we read that $\Delta_Q = 0$ for all CS values.

\subsubsection*{$\SL(2,\F_q)$ Dijkgraaf-Witten theory}

In the finite setting, we expect the existence of flat connections to depend on various divisibility conditions between $n \pm 2$ and $q \pm 1$. Indeed, let us define the quantities
\begin{equation} \label{eq:gcdsTnS}
    \begin{aligned}
        & \kappa = \gcd(q-1, n-2), \qquad \kappa' = \gcd(q-1, n+2), \\
        & \tilde{\kappa} = \gcd(q+1, n-2), \qquad \tilde{\kappa}' = \gcd(q+1, n+2).
    \end{aligned}
\end{equation}
We show the possible solutions to the fundamental group relations in Table \ref{tab:torusTnSsl2q}. All abelian flat connections have their analog in the complex case whenever both $n-2$ and $n+2$ divide $q^2-1$. As for the upper triangular ones, their total number (either 0, 2 or 4) will depend on the specific values of $n$ and $q$, which govern how many solutions exist for the equation $\alpha^2 + \alpha^{-2} = n$ in $\F_q^\times$. In contrast, in the complex case there always exist 4 distinct (and real) solutions.
\begin{table}[ht!]
    \centering
    \begin{tabular}{Sc | Sl}
         Conjugacy class of $X$ & Possible solutions \\
         \hline
        $X = \pm \I$ & If $X = \I$:\, $Y = \I, \quad Z \in \SL(2,\F_q)$ \\
        \hline
        $X = d(\lambda)$ & If $\kappa \neq 1,\, \lambda = \theta^{\ell \frac{q-1}{\kappa}}$:\, $Y = X, \quad Z = d(*)$ \\
         & If $\kappa' \neq 1,\, \lambda = \theta^{\ell \frac{q-1}{\kappa'}}$:\, $Y = X^{-1}, \quad Z = \mqty(0 & * \\ -*^{-1} & 0)$ \\
        \hline
        $X = d'(\xi)$ & If $\tilde{\kappa} \neq 1,\, \xi = \vartheta^{\ell \frac{q^2-1}{\tilde{\kappa}}}$:\, $Y = X, \quad Z = d'(*)$ \\
         & If $\tilde{\kappa}' \neq 1,\, \xi = \vartheta^{\ell \frac{q^2-1}{\tilde{\kappa}'}}$:\, $Y = X^{-1}, \quad Z = A \mqty(0 & \eta \\ -\eta^{-1} & 0) A^{-1}$ \\
        \hline
        $X = \epsilon u(1,a)$ & If $\epsilon = 1$:\, $Y = u(1,a \alpha^{-2}),\quad Z = u(\alpha, *)$,\ with $\alpha^2 + \alpha^{-2} = n$
    \end{tabular}
    \caption{Moduli space of $\SL(2,\F_q)$ connections on $\Sigma_{T^nS}$. Here $A$ is a matrix in $\SL(2,\F_{q^2})$ diagonalizing $d'(\xi)$, and $\eta \in \F^\times_{q^2}$ such that $Z \in \SL(2,\F_q)$.}
    \label{tab:torusTnSsl2q}
\end{table}

We now compute the finite field partition function by taking the trace of the gluing matrix $T^nS$:
\begin{multline}
    Z_\DW(\Sigma_{T^nS}) =  1 + \sum_{\ell = 1}^{\frac{\kappa -1}{2}} e^{-2\pi i \frac{n-2}{\kappa} \frac{12 \ell^2}{\kappa}} + \sum_{\ell' = 1}^{\frac{\kappa' -1}{2}} e^{-2\pi i \frac{n+2}{\kappa'}\frac{12 \ell'^2}{\kappa'}}  \\ + \sum_{\tilde{\ell} = 1}^{\frac{\tilde{\kappa} -1}{2}} e^{-2\pi i \frac{n-2}{\tilde{\kappa}} \frac{12 \tilde{\ell}^2}{\tilde{\kappa}}} + \sum_{\tilde{\ell}' = 1}^{\frac{\tilde{\kappa}' -1}{2}} e^{-2\pi i \frac{n+2}{\tilde{\kappa}'} \frac{12 \tilde{\ell}'^2}{\tilde{\kappa}'}} + |\{\alpha \in \F_q^\times \mid \alpha^2 + \alpha^{-2} = n\}|.
\end{multline}
This expression exactly agrees with our conjecture in the case where both $n \pm 2$ divide $q^2-1$, as all factors defined in Eq.\ (\ref{eq:gcdsTnS}) equal $n \pm 2$ accordingly. Otherwise, the roots of unity corresponding to each flat connection $\gamma$ have order a divisor of $o(\gamma)$ instead. Of course, this is to be seen as a slight generalization of our conjecture to the case where $o(\gamma)$ does not exactly divide $q^2-1$. In any case, such divisibility conditions do not change the leading factors of $q$ corresponding to each Chern-Simons value --- all of them go as $T_Q(q) \sim 1$, as expected from the continuous case.

\subsection{Seifert fibered homology spheres}

\label{sec:seifertHS}

Seifert fibered homology spheres\footnote{The interested reader can find more details in~\cite{saveliev2002invariants}.} over $S^2$ constitute a rich family of homology spheres parameterized by a list of pairwise coprime integers. The Poincaré homology sphere and the Brieskorn homology spheres are particular cases of these manifolds. 

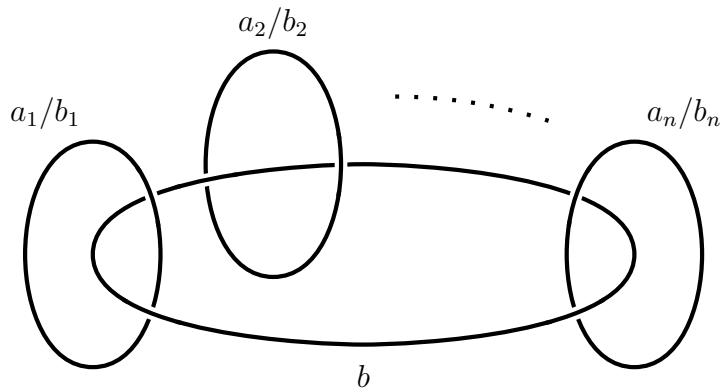
\begin{figure}[ht!]
    \centering
    \begin{tikzpicture}[scale=0.6]
        \begin{knot}[end tolerance=1pt]
        \strand[ultra thick] (-6, 0) 
          .. controls ++(90:1.8) and ++(0:-1) .. (0,2) 
          .. controls ++(0:1) and ++(90:1.8) .. 
          (6,0)
          .. controls ++(90:-1.8) and ++(0:1) .. (0,-2) node[below=3] {$b$}
          .. controls ++(0:-1) and ++(90:-1.8) .. (-6, 0);
        
        \strand[ultra thick] (-7.5, 0) 
          .. controls ++(90:-1.2) and ++(0:-1) .. (-6,-2.5)
          .. controls ++(0:1) and ++(270:1.2) .. (-4.5,0)
          .. controls ++(270:-1.2) and ++(0:1) .. (-6,2.5) node[above left] {${a_1}/{b_1}$}
          .. controls ++(180:1) and ++(90:1.2) .. (-7.5,0)
        ;

        \strand[ultra thick] (-7.5+4, 0+2) 
          .. controls ++(90:-1.2) and ++(0:-1) .. (-6+4,-2.5+2)
          .. controls ++(0:1) and ++(270:1.2) .. (-4.5+4,0+2)
          .. controls ++(270:-1.2) and ++(0:1) .. (-6+4,2.5+2) node[above] {${a_2}/{b_2}$}
          .. controls ++(180:1) and ++(90:1.2) .. (-7.5+4,0+2)
        ;
        
        \strand[ultra thick] (-7.5+12, 0) 
          .. controls ++(90:-1.2) and ++(0:-1) .. (-6+12,-2.5)
          .. controls ++(0:1) and ++(270:1.2) .. (-4.5+12,0)
          .. controls ++(270:-1.2) and ++(0:1) .. (-6+12,2.5) node[above right] {${a_n}/{b_n}$}
          .. controls ++(180:1) and ++(90:1.2) .. (-7.5+12,0)
        ;
        
        \draw[ultra thick, dotted, dash pattern= on 1.5pt off 5.7pt]  (0.7,3.5) arc (90:72:11);
        
        \flipcrossings{1,4,5}
        \end{knot}
    \end{tikzpicture}
    \caption{A surgery presentation of the manifold $M(b;(a_1,b_1),\dots,(a_n,b_n))$.}
    \label{fig:seifertLink}
\end{figure}

A Seifert fibered space over $S^2$ is denoted by $M =M(b;(a_1,b_1),\dots,(a_n,b_n))$. The data $(b;(a_1,b_1),\dots,(a_n,b_n))$ are called the Seifert invariants of $M$, while $n$ is the number of exceptional fibers.
The three-manifolds $M$ can be obtained by performing rational Dehn surgery on the link in Figure~\ref{fig:seifertLink}.

Such manifolds are not in general homology spheres, but one can check that $H_1(M,\mathbb{Z})$ vanishes if and only if
\begin{equation}
     \sum_{i=1}^n \frac{b_i}{a_i} = \frac{1}{a_1 \cdots a_n} +b.
\end{equation}
This implies that the $a_i$ are relatively prime pairwise and the $b_i$ are determined modulo $a_i$. Moreover, $b$ can be set to zero without loss of generality. Thus, the previous condition together with the values of $a_i$, uniquely determines the homology $3$-sphere $M =\Sigma(a_1, \dots, a_n)$.
Notice that we can always restrict $a_2, \dots, a_n$ to be odd, so that $b_2, \dots, b_n$ can be chosen to be even. Furthermore, the fundamental group has the following presentation:
\begin{equation}
    \pi_1(\Sigma(a_1,\dots, a_n)) = \langle x_1, \dots, x_n, h \mid x_1 \cdots x_n =1 , \,  x_i^{a_i}h^{-b_i} =1 , \, [x_i, h] =1, \,i=1, \dots, n\rangle,
\end{equation}
which can be deduced from the surgery presentation.

Much is known about the $\SU(2)$ Chern-Simons theory on Seifert homology spheres. It was shown in~\cite{fintushel1990instanton} that the moduli space of $\SU(2)$ flat connections consists of even-dimensional components of dimension at most $2n-6$. The $\SL(2,\C)$ theory has also been studied; in~\cite{andersen2022resurgence} it was shown that the set of connected components of the moduli space of non-trivial $\SL(2,\C)$-flat connections $\mathcal{M}^*(M,\SL(2,\C))$ is one-to-one with the set $L(a_1,\dots,a_n)$ defined as follows. Its elements are $n$-tuples of at least three non-vanishing integers $\vec{l}=(l_1, \dots, l_n)$ such that $0\le l_1 \le a_1$ and $0\le l_i \le (a_i -1) /2$ for all $i=2,\dots,n$. Let us give a quick idea of how this correspondence works, as the same line of reasoning will be repeated when looking at the case over finite fields. 
For each $\vec{l} \in L(a_1, \dots,a_n)$ one can show that there exists a non-trivial representation $\gamma_{\vec{l}} : \pi_1(M)\rightarrow \SL(2,\C)$ such that
\begin{equation}
    \begin{gathered}
        \gamma_{\vec{l}}(h) = (- \I)^{l_1}, \qquad
        \gamma_{\vec{l}}(x_1)  \in \left[ d(e^{\pi i \, l_1 /a_1}) \right], \qquad
        \gamma_{\vec{l}}(x_i)  \in \left[ d(e^{2\pi i \,l_i /a_i})\right]  \ \text{for $i \ne 1$},
    \end{gathered}
\end{equation}
where square brackets denote the conjugacy class of a matrix in $\SL(2,\C)$. The opposite implication is straightforward once we notice that $M$ being an integer homology sphere forces the image of $h$ to lie in the center of $\SL(2,\C)$.

The set $L(a_1,\dots,a_n)$ also characterizes the values of the Chern-Simons action. Indeed, a further result of~\cite{andersen2022resurgence} states that
\begin{equation}
\CS[\gamma_{\vec{l}} \,] = - \frac{a_1 \cdots a_n}{4}\left(\frac{l_1}{a_1}+ 2\sum_{i=2}^n \frac{l_i}{a_i}\right)^2 \mod 1
\label{eqn:andersenCSvalue}
\end{equation}
where $\gamma_{\vec{l}}$ is the flat connection corresponding to $\vec{l}$. We remark that the computation of the CS functional is based on the method introduced by Kirk and Klassen in~\cite{KirkKlassen90}, generalized to $\SL(2,\C)$-flat connections. Furthermore, the leading order of $k$ in the asymptotic expansion can be obtained by studying the connected components of flat connections in $\mathcal{M}^*(M,\SL(2,\C))$. In particular, their dimension is $2n  -6 -2m$ where $m$ is the number of zero components of the vector $\vec{l}$ corresponding to a flat connection in that connected component.

\subsubsection{Seifert spaces with three exceptional fibers}

For simplicity, we restrict to the case of three exceptional fibers. The manifold $\Sigma(a_1,a_2,a_3)$ is also known as a Brieskorn homology sphere, which was originally defined as the intersection between the unit sphere in $\C^3$ and the complex algebraic surface $z_1^{a_1}+z_2^{a_2}+z_3^{a_3}=0$.

\subsubsection*{Asymptotic expansion of $\SL(2,\C)$ Chern-Simons theory}

The  family  of  Brieskorn   homology  spheres is  very  special since the  moduli  space  of non-trivial flat connections $\mathcal{M}^*(\Sigma(a_1, a_2, a_3), \SL(2,\C))$  is finite, with  cardinality  given  by  the  $\SL(2, \C)$  Casson invariant  introduced  by  Curtis~\cite{curtis2001intersection, curtis2003erratum}: 
\begin{equation}
    \lambda_{\SL(2,\C)}(M) = \frac{(a_1 -1)(a_2-1)(a_3-1)}{4}.
\end{equation}
For reference, in~\cite{fintushel1990instanton} Fintushel and Stern describe a combinatorial formula for the cardinality of the $\SU(2)$ moduli space. Consider triples of integers $(n_1,n_2,n_3) \in \Z^3$ such that $0 < n_i < a_i$ for $i=1,2,3$ and introduce the quantities
\begin{align}
    \sigma_+ &= \left|\left\{(n_1,n_2,n_3) : 0 <\frac{n_1}{a_1}+\frac{n_2}{a_2}+\frac{n_3}{a_3} < 1 \ (\text{mod } 2) \right\}\right|, \\
    \sigma_- &= \left|\left\{(n_1,n_2,n_3) : -1 <\frac{n_1}{a_1}+\frac{n_2}{a_2}+\frac{n_3}{a_3} < 0 \ (\text{mod } 2) \right\}\right|.
\end{align}
Then, the cardinality of $\mathcal{M}^*(\Sigma(a_1,a_2,a_3),\SU(2))$ is given by $(\sigma_- - \sigma_+) /4$.

We can apply this formula in two simple examples. For the Poincaré homology sphere, i.e.\ $M =\Sigma(2,3,5)$, we find that both $\mathcal{M}^*(M,\SL(2,\C))$ and $\mathcal{M}^*(M,\SU(2))$ contain two points. However, the first non-trivial example is $M=\Sigma(2,3,7)$, which still has only two non-trivial inequivalent $\SU(2)$ flat connections, while it has an additional point in $\mathcal{M}^*(M,\SL(2,\C))$, corresponding to an orbit of $\SL(2,\R)$ flat connections as argued in~\cite{andersen2022resurgence}. 
Thus, this example provides a great benchmark to test whether the $\SL(2,\F_q)$ DW theory detects this additional contribution.

Thus, we have all the ingredients to compare with the asymptotic expansion of the $\SL(2,\C)$ CS partition function on a Brieskorn homology sphere $\Sigma(a_1,a_2,a_3)$. Indeed, the value of the action on critical points is easily evaluated from~\eqref{eqn:andersenCSvalue}. Moreover, the leading order of $k$ is $k^{-3/2}$ for the trivial connection and $k^0$ for the other flat connections, see~\cite{FreedGompf91}.

\subsubsection*{$\SL(2,\F_q)$ Dijkgraaf-Witten theory}

Consider now the DW theory on Brieskorn homology spheres $\Sigma(a_1,a_2,a_3)$. Suppose for simplicity that each $a_i$ is prime.
While we will not use the lattice gauge theory description of Dijkgraaf-Witten theory, we still believe that it is interesting to study the moduli space of flat connections $\Hom(\pi_1(\Sigma(a_1,a_2,a_3)),\SL(2,\F_q))$. Remarkably, the same ideas that appeared in~\cite{andersen2022resurgence} and were presented above for the $\SL(2,\C)$ case can be generalized to $\SL(2,\F_q)$ with a bit of caution. Then, it is not too hard to prove that
\begin{multline}
    |\Hom(\pi_1(\Sigma(a_1,a_2,a_3)),\SL(2,\F_q))| =\\
    1 + \frac{(\gcd(a_1,q^2-1)-1)(\gcd(a_2,q^2-1)-1)(\gcd(a_3,q^2-1)-1)}{4}|\SL(2,\F_q)|  
\end{multline}
away from special primes, i.e.\ $q$ must not be a power of any prime number in the decomposition of $a_1a_2a_3$. Notice that if $a_1a_2a_3 \mid q^2-1$ then the factor multiplying $|\SL(2,\F_q)|$ is equal to the $\SL(2,\C)$ Casson invariant $\lambda_{\SL(2,\C)}$.

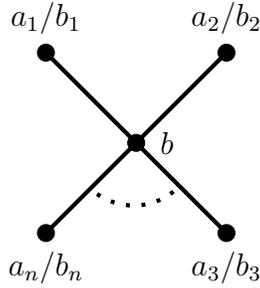
\begin{figure}[ht!]
    \centering
    \begin{tikzpicture}[scale=0.4]
        \draw[ultra thick] (0,0)  -- (3,3);
        \draw[ultra thick] (0,0)  -- (-3,3);
        \draw[ultra thick] (0,0) -- (3,-3);
        \draw[ultra thick] (0,0) -- (-3,-3);
        \draw[ultra thick, dotted, dash pattern= on 1.5pt off 4.4pt]  (-1.5+0.2,-1.5-0.2) arc (180+55:360-50:2.12);
        \filldraw[black] (3,-3) circle (8pt) node[below=3] {$a_3/b_3$};
        \filldraw[black] (3,3) circle (8pt) node[above=3] {$a_2/b_2$};
        \filldraw[black] (-3,3) circle (8pt) node[above=3] {$a_1/b_1$};
        \filldraw[black] (-3,-3) circle (8pt) node[below=3] {$a_n/b_n$};
        \filldraw[black] (0,0) circle (8pt) node[right=5] {$b$};
    \end{tikzpicture}
    \caption{The plumbing graph of $M(b;(a_1,b_1),\dots,(a_n,b_n))$.}
    \label{fig:seifertPlumbingGraph}
\end{figure}

Brieskorn homology spheres, and actually all Seifert fibrations over $S^2$, are plumbed manifolds with star-shaped plumbing graphs. For example, the plumbing graph for the manifold $M(b;(a_1,b_1),\dots,(a_n,b_n))$ can be easily deduced from Figure~\ref{fig:seifertLink} and it is shown in Figure~\ref{fig:seifertPlumbingGraph}. 
Here, we admit vertices with rational weights, which have to be understood as a sequence of $l$ vertices with integer weights $m_1, \dots, m_l$ where $[m_1, \dots, m_l]$ is the continued fraction expression of the rational weight, as shown in Figure~\ref{fig:rationalWeights}. Recall that given a rational number $a/b$, we say that $[m_1, \dots, m_l]$ is a continued fraction for it if
\begin{equation}
    \frac{a}{b} =  m_l - \frac{1}{m_{l-1}- \frac{1}{\dots - \frac{1}{m_1}}}.
\end{equation}
The direct consequence of this fact is that we can compute the partition function of any Brieskorn homology sphere by simply applying the formula in Equation~\eqref{eqn:plumbedDWPartitionFunction}.
\begin{figure}[ht!]
    \centering
    \begin{tikzpicture}[scale=0.4]
        \draw[ultra thick,dotted] (-3,0)  -- (-2,0);
        \draw[ultra thick] (-2,0)  -- (2,0);
        \filldraw[black] (2,0) circle (8pt) node[below=3] {$a/b$};
        \node at (6,0) {$\equiv$};
        \draw[ultra thick,dotted] (10,0)  -- (11,0);
        \draw[ultra thick] (11,0)  -- (15,0);
        \filldraw[black] (15,0) circle (8pt) node[below=3] {$m_l$};
        \draw[ultra thick] (15,0)  -- (20,0);
        \filldraw[black] (20,0) circle (8pt) node[below=3] {$m_{l-1}$};
        \draw[ultra thick,dotted, dash pattern=on 1.5pt off 6pt] (21,0)  -- (24,0);
        \filldraw[black] (25,0) circle (8pt) node[below=3] {$m_2$};
        \draw[ultra thick] (25,0)  -- (30,0);
        \filldraw[black] (30,0) circle (8pt) node[below=3] {$m_1$};
    \end{tikzpicture}
    \caption{When the weight of a vertex is rational, the vertex gets replaced with a sequence of vertices corresponding to its continued fraction expression.}
    \label{fig:rationalWeights}
\end{figure}
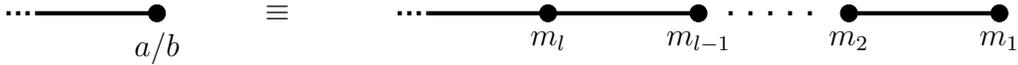
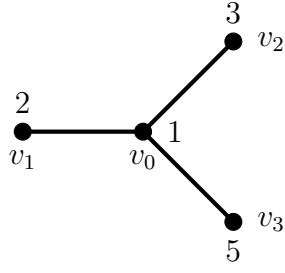
\begin{figure}[ht!]
    \centering
    \begin{tikzpicture}[scale=0.4]
        \draw[ultra thick] (0,0)  -- (3,3);
        \draw[ultra thick] (0,0) -- (3,-3);
        \draw[ultra thick] (0,0) -- (-4,0);
        \filldraw[black] (3,-3) circle (8pt) node[below=3] {$5$}node[right=5]{$v_3$};
        \filldraw[black] (3,3) circle (8pt) node[above=3] {$3$} node[right=5]{$v_2$};
        \filldraw[black] (-4,0) circle (8pt) node[above=3] {$2$} node[below=3]{$v_1$};
        \filldraw[black] (0,0) circle (8pt) node[right=5] {$1$} node[below=3]{$v_0$};
    \end{tikzpicture}
    \caption{The plumbing graph of the Poincaré homology sphere. We denote by $s_i$ the state corresponding to the vertex $v_i$.}
    \label{fig:poincarePlumbingGraph}
\end{figure}

Let us consider the Poincaré homology sphere $\Sigma(2,3,5)$ as an example of how to carry on the computation of the partition function. Its plumbing graph is shown in Figure~\ref{fig:poincarePlumbingGraph} and it allows us to find
\begin{equation}
    Z_\DW(\Sigma(2,3,5)) = \sum_{s_0,s_1,s_2,s_3} S_{s_0,s_1} S_{s_0,s_2} S_{s_0,s_3} S_{0,s_0}^{-1} S_{0,s_1} S_{0,s_2} S_{0,s_3} T_{s_0,s_0} T_{s_1,s_1}^2 T_{s_2,s_2}^3 T_{s_3,s_3}^5.
\end{equation}
Being a homology sphere, the state corresponding to the trivalent vertex is forced to be of the form $s_0 =(\epsilon\I, \chi)$, $\epsilon$ being a sign. This is in agreement with the fact that the image of the generator $h \in \pi_1(\Sigma(2,3,5))$ must belong to the center of $\SL(2,\F_q)$. Then, if we remove the trivial contribution corresponding to having all the $s_i$ of the form $(\pm \I, \chi)$, we find that $s_1,s_2$ and $s_3$ have to be the form $(a,\psi)$, where $a$ is a diagonal or codiagonal conjugacy class representative, and $\psi$ is a character of $T$ or $T'$, respectively (notice that this is what fails when $p =2,3,5$). Hence, the partition function can be written as:
\begin{multline}
Z_{\DW}(\Sigma(2,3,5)) = \frac{1}{|\SL(2,\F_q)|} +
    \frac{|\SL(2,\F_q)|}{q_{\pm}^6} \sum_{\epsilon = \pm} \sum_{a_1,a_2,a_3} \varepsilon_{a_1}(a_1)^2 \, \varepsilon_{a_2}(a_2)^{3} \,\varepsilon_{a_3}(a_3)^{5} \times \\
    \sum_{\chi} \frac{\chi(a_1) \chi(a_2) \chi(a_3) \chi(\epsilon \I)}{\chi(\I)^2}
    \sum_{\psi_1,\psi_2,\psi_3}\psi_1(\epsilon a_1^2) \, \psi_2(\epsilon a_2^3) \,  \psi_3(\epsilon a_3^5),
\end{multline}
where $q_{\pm}$ is either $q+1$ or $q-1$.
The sum over $\psi_1$ forces $a_1$ to satisfy $a_1^2 = \epsilon \I$, however the case $\epsilon = 1$ is ruled out since $a_1 \ne \I,-\I$. We get similar conditions for the other $a_i$. Then, we obtain:
\begin{multline}
\label{eqn:phsPFpart1}
Z_{\DW}(\Sigma(2,3,5)) =\frac{1}{|\SL(2,\F_q)|} +
    \frac{|\SL(2,\F_q)|}{q_{\pm}^3} \sum_{a_1,a_2,a_3} \varepsilon_{a_1}(a_1)^2 \, \varepsilon_{a_2}(a_2)^{3} \,\varepsilon_{a_3}(a_3)^{5} \times \\
    \delta(a_1^2 = -\I) \delta(a_2^3 = -\I  ) \delta(a_3^5 = -\I)
    \sum_{\chi} \frac{\chi(a_1) \chi(a_2) \chi(a_3) \chi(-\I)}{\chi(\I)^2}.
\end{multline}
The evaluation of the sum over the $\SL(2,\F_q)$ characters is quite tricky and we discuss it in Appendix~\ref{app:sums}. There we prove that
\begin{equation}
    \label{eqn:PHSsumOverIrrChar}
    \frac{|\SL(2,\F_q)|}{q_{\pm}^3} \sum_{\chi} \frac{\chi(a_1) \chi(a_2) \chi(a_3) \chi(-\I)}{\chi(\I)^2} = 1.
\end{equation}
Plugging this result into the expression above, the partition function for the Poincaré homology sphere is
\begin{equation}
Z_{\DW}(\Sigma(2,3,5)) =\frac{1}{|\SL(2,\F_q)|} +
     (e^{2\pi i/10} + e^{-2\pi i/10}) \delta(5 |q^2-1).
\end{equation}

We now consider the general case of the Brieskorn homology sphere $\Sigma(a_1,a_2,a_3)$, assuming only that the $a_i$ are prime. This implicitly fixes $b_1,b_2,b_3$ to satisfy $b_1/a_1+b_2/a_2+b_3/a_3=1/(a_1a_2a_3)$, which makes the manifold a homology sphere. For simplicity, we will only consider the case of $a_1a_2a_3 \mid q^2-1$. A similar computation as in the case of the Poincaré homology sphere yields
\begin{multline}
    Z_\DW(\Sigma(a_1,a_2,a_3)) = \\ \frac{1}{|\SL(2,\F_q)|} + \sum_{k_1,k_2,k_3} e^{-6\pi i\qty(\frac{b_1}{a_1}k_1^2 + \frac{b_2}{a_2}k_2^2 +\frac{b_3}{a_3}k_3^2)}+ \sum_{m_1,m_2,m_3} e^{-6\pi i\qty(\frac{b_1}{a_1}m_1^2 + \frac{b_2}{a_2}m_2^2 +\frac{b_3}{a_3}m_3^2)},
\end{multline}
where the first sum is over odd $k_i$ with $1 \le k_1 \le a_1-1$ and $1 \le k_i \le a_i-2$ for $i=2,3$, while the second sum is over even $m_i$ with $2 \le m_i \le a_i -1$ for $i=1,2,3$. Notice that when $a_1 = 2$, the second sum does not contribute.

The phases appearing in the DW partition function have to be compared with the values of the CS functional in~\eqref{eqn:andersenCSvalue}, which can be written as:
\begin{equation}
    12\CS(l_1,l_2,l_3) =-3a_1a_2a_3\left( \frac{l_1}{a_1} + \frac{2l_2}{a_2} + \frac{2l_3}{a_3} \right)^2 \mod 1, 
\end{equation}
with $\vec{l} \in L(a_1,a_2,a_3)$. The correspondence between these two sets of values is not evident, but it is easy to check. Hence, this result is in agreement with the statement of our conjecture, since the leading power of $q$ also matches with $-2\Delta_Q$.

\subsubsection{A simple example with four exceptional fibers}

\begin{figure}[ht!]
    \centering
    \begin{tikzpicture}[scale=0.4]
        \draw[ultra thick] (0,0)  -- (3,3);
        \draw[ultra thick] (0,0)  -- (-3,3);
        \draw[ultra thick] (0,0) -- (3,-3);
        \draw[ultra thick] (0,0) -- (-3,-3);
        \filldraw[black] (3,-3) circle (8pt) node[below=3] {$11$};
        \filldraw[black] (3,3) circle (8pt) node[above=3] {$3$};
        \filldraw[black] (-3,3) circle (8pt) node[above=3] {$2$};
        \filldraw[black] (-3,-3) circle (8pt) node[below=3] {$13$};
        \filldraw[black] (0,0) circle (8pt) node[right=5] {$-1$};
    \end{tikzpicture}
    \caption{The plumbing graph of $\Sigma(2,3,11,13)$.}
    \label{fig:seifert4PlumbingGraph}
\end{figure}
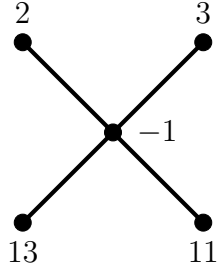

For completeness, we consider a simple case of a Seifert homology sphere with four fibers, namely $\Sigma(2,3,11,13)$ with plumbing graph as in Figure~\ref{fig:seifert4PlumbingGraph}. In the case of four exceptional fibers, the moduli space of flat connections contains connected components of dimension two, in addition to isolated points. Then, in the DW partition function we expect contributions of order $q^2$ from these flat connections.

The computation proceeds as in the case of three exceptional fibers with the exception that up to one of the four generators $x_i \in \pi_1(\Sigma(2,3,11,13))$ can be chosen to be sent to $\pm \I$. Then, we obtain the following partition function:
\begin{multline}
    Z_\DW(\Sigma(2,3,11,13)) = \frac{1}{|\SL(2,\F_q)|}+2\sum_{l_3,l_4} e^{-6\pi i\qty(\frac{l_3^2}{11}+\frac{l^2_4}{13})} + \sum_{l_3,l_4} e^{-6\pi i\qty(\frac{1}{2}+\frac{l_3^2}{11}+\frac{l^2_4}{13})} + \\
    \sum_{l_4} e^{-6\pi i\qty(\frac{1}{2}+\frac{l^2_4}{13})}+ \sum_{l_3} e^{-6\pi i\qty(\frac{1}{2}+\frac{l_3^2}{11})} + (q^2+1) \sum_{l_3,l_4} e^{-6\pi i\qty(\frac{1}{2}+\frac{l_3^2}{11}+\frac{l^2_4}{13})},
\end{multline}
where the sums are over odd $l_3$ and $l_4$ with $1 \le l_3 \le 9$ and $1 \le l_4 \le 11$. The phases are again in agreement with $12\CS$ as one can check via~\eqref{eqn:andersenCSvalue}. Moreover, the leading order of $q$ in the prefactors of each contribution also matches with $-2\Delta_Q$ as expected.

\subsubsection{Homology spheres that are not Seifert spaces} \label{sec:noGlobal}

More general homology spheres may be constructed from plumbing graphs satisfying certain conditions --- from~\eqref{eqn:homologyPlumbed} it is evident that $M_\Gamma$ is an integer homology sphere if and only if all the $g_i = 0$ and $\det B_\Gamma = \pm 1$.

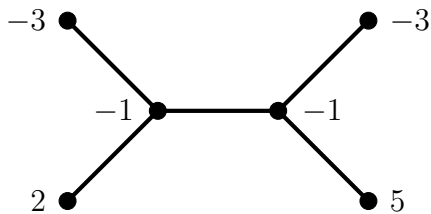
\begin{figure}[ht!]
    \centering
  \begin{tikzpicture}[scale=0.4]
\draw[ultra thick] (5,0)  -- (9,0);
  \draw[ultra thick] (9,0) -- (12,3);
   \draw[ultra thick] (9,0) -- (12,-3);
    \draw[ultra thick] (5,0) -- (2,-3);
     \draw[ultra thick] (5,0) -- (2,3);
    \filldraw[black] (5,0) circle (8pt) node[left=5] {$-1$};
    \filldraw[black] (2,3) circle (8pt) node[left=4] {$-3$};
    \filldraw[black] (2,-3) circle (8pt) node[left=4] {$2$};
    \filldraw[black] (9,0) circle (8pt) node[right=5] {$-1$};
    \filldraw[black] (12,3) circle (8pt) node[right=4] {$-3$};
    \filldraw[black] (12,-3) circle (8pt) node[right=4] {$5$};
\end{tikzpicture}
    \caption{An example of a plumbed integer homology sphere that is not a Seifert manifold. Note its plumbing graph $\Gamma_1$ is not star-shaped.}
    \label{fig:first-plumbed-manifold}
\end{figure}

Let us now discuss a concrete example. Consider the plumbing graph $\Gamma_1$ shown in Figure~\ref{fig:first-plumbed-manifold}. It is immediate to check that it is an integer homology sphere. Furthermore, it cannot be realized as a Seifert fibered space as its plumbing graph is in normal form and is not star-shaped, see~\cite{Neumann81}. Thus, it is an example of a manifold that does not admit a global geometry in the sense of Thurston's geometrization conjecture. 

The possible values of the Chern-Simons functional on flat connections were determined in \cite{GukovPutrov24} through a \emph{finite-dimensional model} of analytically continued $\SU(2)$ Chern-Simons theory. Here we list their results, organized into four different kinds of contributions:
\begin{itemize}
    \item One contribution comes from the trivial flat connection, and has the usual leading order $\Delta_Q = 3/2$ as the manifold is a homology sphere.
    \item The second kind consists of flat connections such that $12\CS \in \{ 1/26, 3/26, 9/26, \allowbreak  17/26,23/26,25/26 \}$ , with leading order in $k$ given by $\Delta_Q = 0 $.
    \item The third kind consists of flat connections such that $12\CS \in \{ 6/35, 19/35, 24/35, \allowbreak26/35, 31/35, 34/35\}$, with leading order in $k$ given by $\Delta_Q = 0$.
    \item Finally, the fourth kind is such that $12\CS \in \{ 1/10,9/10\}$, and has leading order in $k$ given by $\Delta_Q = -1/2$.
\end{itemize}
Remarkably, the last kind of contribution is part of a phenomenon that was first noticed in~\cite{GukovPutrov24}, namely the existence of the so-called \emph{flat connections at infinity}.
A flat connection at infinity is a discrete (or continuous) family of $\SL(2,\C)$ connections $\{ \gamma_\varepsilon\}$ which is singular in the limit $\varepsilon\to\infty$, but the curvature tends to 0 and $\CS[\gamma_\varepsilon]$ has a finite limit $\CS[\gamma_\infty]$. 
In this work, the authors study some necessary conditions for their existence on manifolds obtained as $\pm1$-surgery on knots. It can be proved that the fourth kind of contribution actually comes from flat connections at infinity, by explicitly constructing the family of $\SL(2,\C)$ flat connections $\{\gamma_\varepsilon\}$.

On the other hand, we can consider the $\SL(2,\F_q)$ Dijkgraaf-Witten partition function. If we suppose that $5\cdot7\cdot13 \mid q^2-1$, Equation~\eqref{eqn:plumbedDWPartitionFunction} yields
\begin{equation}
    Z_\DW(M_{\Gamma_1}) = \frac{1}{|\SL(2,\F_q)|} + \sum_{\substack{l=1 \\ \text{odd}}}^{11} e^{2\pi i\qty(\frac{1}{2}-\frac{6}{13}l^2)} +\sum_{\substack{l=1 \\ \text{odd}}}^{5} \sum_{\substack{l'=1 \\ \text{odd}}}^{3} e^{2\pi i(\frac{4}{7}l^2+\frac{2}{5}l'^2)}.
\end{equation}
Notice that this is in agreement with our conjecture for the standard $\SL(2,\C)$ flat connections, but it does not detect flat connections at infinity.

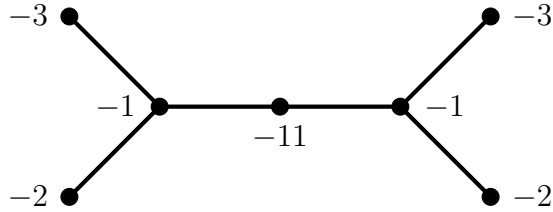
\begin{figure}[ht!]
    \centering
    \begin{tikzpicture}[scale=0.4]
        \draw[ultra thick] (5,0)  -- (9,0);
        \draw[ultra thick] (9,0) -- (13,0);
        \draw[ultra thick] (13,0) -- (16,3);
        \draw[ultra thick] (13,0) -- (16,-3);
        \draw[ultra thick] (5,0) -- (2,-3);
        \draw[ultra thick] (5,0) -- (2,3);
        \filldraw[black] (5,0) circle (8pt) node[left=5] {$-1$};
        \filldraw[black] (2,3) circle (8pt) node[left=4] {$-3$};
        \filldraw[black] (2,-3) circle (8pt) node[left=4] {$-2$};
        \filldraw[black] (9,0) circle (8pt) node[below=3] {$-11$};
        \filldraw[black] (13,0) circle (8pt) node[right=5] {$-1$};
        \filldraw[black] (16,3) circle (8pt) node[right=4] {$-3$};
        \filldraw[black] (16,-3) circle (8pt) node[right=4] {$-2$};
    \end{tikzpicture}
    \caption{The plumbing graph $\Gamma_2$ of the $1$-surgery on the right-handed granny knot $3_1 \# 3_1$.}
    \label{fig:grannyknotplumbing}
\end{figure}

We conclude this section with one last example. Consider the manifold obtained as the $1$-surgery on the right-handed granny knot $3_1 \# 3_1$, whose plumbing graph $\Gamma_2$ is shown in Figure~\ref{fig:grannyknotplumbing}. This example was already considered in~\cite{GukovPutrov24} where it was stated that it admits flat connections at infinity. 

The following results are obtained by applying the finite-dimensional model of~\cite{GukovPutrov24}. The asymptotic expansion gets contributions from:
\begin{itemize}
    \item The trivial flat connection.
    \item A family of flat connections with $12\CS \in \{0, 1/11, 3/11,4/11,5/11,9/11\}$ and $\Delta_Q=-1/2$.
    \item Two additional families with $12\CS \in \{ 1/2,1/10,9/10\}$ and $\Delta_Q=0$.
\end{itemize}
After a detailed analysis, we found out that this manifold actually admits three flat connections at infinity. However, this is not in disagreement with~\cite{GukovPutrov24} since, as the authors claim, only one of these actually contributes to the asymptotic expansion of the Chern-Simons theory.

The moduli spaces of $\SL(2,\C)$ flat connections has been also studied in the context of $\SL(2,\C)$ Floer cohomology of \cite{abouzaid2020sheaf}. In particular, explicit description of the moduli space for surgeries on the granny knot can be found in~\cite{neithalath2023sl}. In particular, for the $+1$-surgery we expect $4$ points and $5$ copies of $\C^\times$, whereas above we listed a total of $6$ points and $6$ copies of $\C^\times$. 
It was already noted in~\cite{GukovPutrov24} that this method does not detect flat connections at infinity. However, we can compare the two results to count the number of such connections.

Finally, we compute the DW partition function using~\eqref{eqn:plumbedDWPartitionFunction}. Assuming that $5\cdot11 \mid q^2 -1,$ we get
\begin{equation}
    Z_\DW(M_{\Gamma_2}) = \frac{1}{|\SL(2,\F_q)|} + (q\pm1) \sum_{l=1}^5 e^{2\pi i \frac{l^2}{11}} + 2 \sum_{l=1,3} e^{2\pi i \qty(\frac{1}{2}+\frac{3}{5}l^2)}, \qquad \text{if $11 \mid q \pm 1$}.
\end{equation}
Notice that the factor $2$ in the last sum corresponds to two families of connections with the same phase factor. This result is again in complete agreement with our conjecture.

\subsection{Higher-genus plumbings} \label{sec:higherGenus}

The simplest example of a higher-genus plumbing is given by a graph $\Gamma$ consisting of one single vertex of genus $g > 1$ and weight $n$. The corresponding 3-manifold $M$ is then a degree $n$-circle bundle over the Riemann surface $\Sigma_g$ of genus $g$. Its fundamental group reads
\begin{equation}
    \pi_1(M) = \expval{a_1, \dots, a_g, b_1, \dots, b_g, h \ \eval \ \prod_{i = 1}^g [a_i,b_i] = h^n, [a_i, h] = 1, [b_i,h] = 1}.
\end{equation}
It is worth noting that these manifolds are all geometric of type $\widetilde{\SL(2,\R)}$. In what follows, we consider $n$ to be an odd prime for simplicity.

The asymptotic behavior of the $\SU(2)_k$ Chern-Simons partition function can be derived, for example, from the results for general Seifert fibered manifolds in \cite{Hansen05}. Indeed, $M$ can be represented as the Seifert manifold with base $\Sigma_g$ and one single exceptional fiber of type $(1,n)$. Alternatively, one can use the finite-dimensional model of \cite{GukovPutrov24}, which also includes $\SL(2,\C)$ flat connections.

\subsubsection*{Asymptotic expansion of $\SL(2,\C)$ Chern-Simons theory}

The set of all possible CS values on flat connections reads
\begin{equation}
    \CS \in \qty{0, \frac{n}{4}} \cup \qty{-\frac{a^2}{n} \ \eval \ a = 1, \dots, \frac{n-1}{2}}.
\end{equation}
The corresponding leading powers of $k$ are given as follows. For the trivial CS value, $\Delta_Q = 3-3g$. The asymptotic analysis also reveals a subleading contribution of order $k^{2g-\frac{3}{2}}$, which can be singled out through the methods of \cite{Hansen05}, or via the finite-dimensional model in \cite{GukovPutrov24} {(in which case it manifests itself as an additional Lefschetz thimble contributing to the term with trivial CS value in the asymptotic expansion)}. For the value $\CS = \frac{1}{4}$ we have $\Delta_Q = 3-3g$ as well, while for the last set of values $\Delta_Q = \frac{1}{2}-g$ instead.

\subsubsection*{$\SL(2,\F_q)$ Dijkgraaf-Witten theory}

We now assume that $n \mid q^2-1$ in order to obtain non-trivial results (hence $n \neq p$). The finite field partition function can be computed via Eq.\ (\ref{eqn:plumbedDWPartitionFunction}) --- in our case, we have
\begin{multline}
        Z_{\DW}(M) = \sum_{s \in \mathcal{H}(\T^2)} T^n_{s, s} S^{2-2g}_{0, s} \\
        = |G|^{2g-2} \qty(1 + q^{2-2g} + \qty(\frac{q-3}{2}+2^{2g-1})(q+1)^{2-2g} + \qty(\frac{q-1}{2}+2^{2g-1})(q-1)^{2-2g}) \\
        + |G|^{2g-2} \times \begin{cases}
        1 + q^{2-2g} + (2^{2g-1}-1)(q+1)^{2-2g} - 2^{2g-1}(q-1)^{2-2g}, & q = 1 \text{ mod 4}, \\ 
        1 + q^{2-2g} + (2^{2g-1}-1)(q-1)^{2-2g} - 2^{2g-1}(q+1)^{2-2g}, & q = 3 \text{ mod 4},
    \end{cases} \\
    + (q \pm 1)^{2g-1} \sum_{a = 1}^{\frac{n-1}{2}} e^{-2 \pi i \frac{12 a^2}{n}}, \qquad n \mid q \pm 1.
\end{multline}
We have written the result in this particular way to highlight the contributions from each of the states $s = (a, \chi)$ coming from different conjugacy classes: the first, second, and third lines correspond to contributions from states with $a = \I$, $a = -\I $, and $a \in T$ or $a \in T'$ respectively (the latter depending on $a \mid q-1$ or $a \mid q+1$).

First of all, we find contributions of leading order $q^{6g-6}$ in the first and second lines, agreeing with the asymptotics of the two components with CS value $0$ and $\frac{p}{4}$ (which is trivial multiplied by 12) as per our conjecture. In the first line we can further identify particular contributions of order $q^{4g-3}$, reminiscent of the subleading contributions of order $k^{2g-\frac{3}{2}}$ to the term with trivial CS value in the asymptotic expansion. Finally, the last line corresponds to the remaining values of the CS invariant, with a coefficient of leading order $q^{2g-1}$ as expected.

\subsection{An example with \texorpdfstring{$\mathbb{H}^2 \times \R$}{H2 x R} geometry} \label{sec:h2xrExample}

In the previous sections, we have considered six out of the eight types of geometric structures. Except for the hyperbolic case, only the geometry $\mathbb{H}^2 \times \R$ is left. A large class of manifolds with this geometric structure is given by Seifert fibered spaces over $S^2$ with negative orbifold Euler characteristic $\chi = 2 -\sum_i (1-1/a_i)$ and vanishing Seifert Euler number $E = b+ \sum_i b_i/a_i$.

\begin{figure}[ht!]
    \centering
    \begin{tikzpicture}[scale=0.4]
        \draw[ultra thick] (0,0)  -- (0.87*5,0.5*5);
        \draw[ultra thick] (0,0) -- (-0.87*5,2.5);
        \draw[ultra thick] (0,0) -- (-0.87*10,5);
        \draw[ultra thick] (0,0) -- (0.87*10,5);
        \draw[ultra thick] (0,0) -- (0,-5);
        \filldraw[black] (0,0) circle (8pt) node[above=3] {$-1$};
        \filldraw[black] (-0.87*5,2.5) circle (8pt) node[above=3] {$-2$};
        \filldraw[black] (0.87*5,2.5) circle (8pt) node[above=3] {$-2$};
        \filldraw[black] (0.87*10,5) circle (8pt) node[right=4] {$2$};
        \filldraw[black] (-0.87*10,5) circle (8pt) node[left=4] {$2$};
        \filldraw[black] (0,-5) circle (8pt) node[below=4] {$-5$};
    \end{tikzpicture}
    \caption{The plumbing graph of $M(-1;(5,-2),(5,-2),(5,-1))$.}
    \label{fig:h2plumbing}
\end{figure}
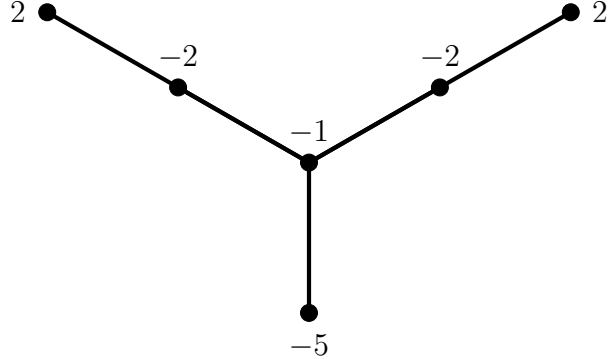

We focus on the manifold $M=M(-1;(5,-2),(5,-2),(5,-1))$, schematically given by the plumbing graph in Figure~\ref{fig:h2plumbing}. Notice that the fundamental group, which can be deduced from the plumbing graph, is given by
\begin{equation}
    \pi_1(M) = \langle x,y,z,h \mid x^5h^2=y^5h^2=z^5h=xyzh=[x,h]=[y,h]=[z,h]=1\rangle,
\end{equation}
and its abelianization gives the first homology group:
\begin{equation}
    H_1(M, \Z) = \Z \times \Z/5\Z.
\end{equation}

A crucial property of manifolds with $E = 0$ is that their linking matrix is singular, as $b_1>0$. Thus, we will not be able to easily compute the asymptotics of the $\SL(2,\C)$ Chern-Simons partition function from the finite-dimensional model of~\cite{GukovPutrov24}.

\subsubsection*{Asymptotic expansion of $\SL(2,\C)$ Chern-Simons theory}

The $\SU(2)$ Chern-Simons partition functions of Seifert fibered spaces are pretty well-known in the literature, see for example~\cite{lawrence1999witten}. Unfortunately, $\SL(2,\C)$ did not get a similar interest. So far, we managed to bypass this issue by employing the finite-dimensional method of~\cite{GukovPutrov24}. However, this fails when the linking matrix is singular. 

In~\cite{Hansen05}, general formulas for the asymptotic expansion of the $\SU(2)$ partition function of Seifert manifolds appear. From these, we can deduce that $\SU(2)$ flat connections contribute with a phase factor given by $\CS \in \{0, 1/5, 4/5\}$ and leading order in $k$ that is $\Delta_Q=0$ for each Chern-Simons value. 
We speculate that the same argument given for $G = \SU(2)$ also applies to $G = \SL(2,\C)$. Then, the genuine $\SL(2,\C)$ flat connections, which do not arise from $\SU(2)$ flat connections, are expected to contribute with $\CS \in \{2/5, 3/5\}$ and $\Delta_Q=0$. We remark that the proof in~\cite{Hansen05} is based on~\cite{auckly1994topological} and on the method introduced by Kirk and Klassen in~\cite{KirkKlassen90}. Generalizations of the latter to $\SL(2,\C)$
already appeared in~\cite{andersen2022resurgence}.

\subsubsection*{$\SL(2,\F_q)$ Dijkgraaf-Witten theory}

In this example, we witness a new peculiar feature of the theory. Already at the untwisted level, there is a qualitative difference between the cases $5\mid q+1$ and $5 \mid q-1$, which is a novelty. Indeed, up to a normalization constant, the untwisted partition function for a generic prime power $q$ is:
\begin{equation}
    |\Hom(\pi_1(M),\SL(2,\F_q)))| = \begin{cases}
        |G|, \qquad &5\nmid q^2-1, \\
        7|G|, \qquad &5 \mid q+1, \\
        31|G|, \qquad &5 \mid q-1.
    \end{cases}
\end{equation}
This feature has consequences even in the twisted theory. Indeed, Equation~\eqref{eqn:plumbedDWPartitionFunction} yields
\begin{align}
    Z_\DW(M) = & \ \frac{2}{|G|} + \frac{2(q^2-1)}{|G|}  + \frac{12}{q\pm 1} (\zeta_5^2 + \zeta_5^{-2}) \\ 
     & +\begin{cases}
        \frac{q-3}{2(q-1)}, \qquad &5 \nmid q-1, \\
        \frac{q-11}{q-1} \cdot \frac{5}{4} (\zeta_5^2 + \zeta_5^{-2}), \qquad &5 \mid q-1
    \end{cases}
    +
    \begin{cases}
        \frac{q-1}{2(q+1)}, \qquad &5 \nmid q+1, \\
        \frac{q-9}{q+1} \cdot \frac{5}{4} (\zeta_5^2 + \zeta_5^{-2}), \qquad &5 \mid q+1
    \end{cases} 
    \notag
    \\
     & +\left( \, \sum_{l_i = 1,2} e^{2\pi i \qty( - \frac{l_1^2}{5} - \frac{l_2^2}{5} + 2 \frac{l_3^2}{5} )}\ + \sum_{l_i = 1,3} e^{2\pi i \qty( \frac{l_1^2}{5} + \frac{l_2^2}{5} -2 \frac{l_3^2}{5} )} \right) \begin{cases}
        \frac{2q-1}{q-1}, \qquad &5 \mid q-1, \\
        \frac{1}{q+1},  \qquad &5 \mid q+1,
    \end{cases} 
    \notag
\end{align}
which results in
\begin{equation}
    Z_{\DW} (M)  = \begin{cases}
        \frac{17q-7}{2(q-1)} + 2(\zeta_5+\zeta_5^{-1}) + \frac{37q - 47}{4(q-1)} (\zeta_5^2 + \zeta_5^{-2}), \qquad &5 \mid q-1, \\
        \frac{q+11}{2(q+1)} + 2(\zeta_5+\zeta_5^{-1}) + \frac{5}{4}\frac{q-1}{q+1} (\zeta_5^2 + \zeta_5^{-2}), \qquad &5 \mid q+1.
    \end{cases} 
\end{equation}
Thus, also this case is in agreement with our conjecture, as $\Delta_Q = 0$ for each value.

\subsection{Hyperbolic manifold counterexample} \label{sec:hyperbolicEx}

\begin{figure}[ht!]
    \centering
    \begin{tikzpicture}[use Hobby shortcut,scale = 1.6]
    \begin{knot}[
    consider self intersections=true,
    ignore endpoint intersections=false,
    flip crossing=4,
    only when rendering/.style={
    }
    ]
    \strand[ultra thick] ([closed]0,0) .. (1.5,1) .. (.5,2) .. (-.5,1) .. (.5,0) ..      (0,-.5) .. (-.5,0) .. (.5,1) .. (-.5,2) .. (-1.5,1) .. (0,0);
    \end{knot}
    \path (0,-.7);
    \end{tikzpicture}
    \caption{The figure eight knot.}
    \label{fig:figure8}
\end{figure}
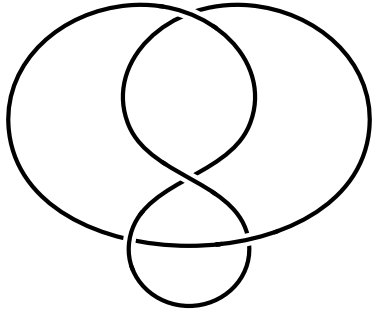

At last, we consider an example --- or even better, a counterexample --- of a hyperbolic 3-manifold. In particular, we study the $-1/2$ surgery on the figure-eight knot $4_1$, depicted in Figure~\ref{fig:figure8}. 
Such a manifold is a closed, hyperbolic, integer homology sphere $M =S_{-1/2}(4_1)$. 

We remark that plumbed manifolds are not hyperbolic, meaning that their geometric decomposition does not contain a hyperbolic component. Furthermore, the partition function of hyperbolic manifolds is not computed via $S$ and $T$ matrices as in~\eqref{eqn:plumbedDWPartitionFunction}. For how to deal with this problem, we refer back to the discussion at the beginning of Section~\ref{sec:DWsl2q}. In short, we fix a triangulation of $S_{-1/2}(4_1)$ and bypass the necessity for a branched triangulation by choosing a cocycle representing $\omega = 24\beta^{-1}c_2(\rho)$ that satisfies~\eqref{eq:cocyclegauge}. Again, that the coefficient of $24$ is crucial as explained in Section~\ref{sec:DWsl2q} and in Appendix~\ref{app:acocycles}.

\subsubsection*{Asymptotic expansion of $\SL(2,\C)$ Chern-Simons theory}

Let us study the moduli space of $\SL(2,\C)$-flat connections. In order to do so, consider the presentation of the fundamental group of the manifold:
\begin{equation}
    \pi_1(S_{-1/2}(4_1)) = \langle x,y \mid x=\omega^{-1}\nu\omega^{-1}\nu , \, y = \nu\omega^{-1}\nu\omega^{-1}\rangle,
\end{equation}
where $\omega = [x^{-1},y]$ and $\nu = [x,y^{-1}]$ are such that $\omega^{-1} \nu$ is a canonical choice for the longitude of the knot complement $S^3 \setminus 4_1$. Non-trivial elements of $\Hom(\pi_1(S_{-1/2}(4_1)),\SL(2,\C))$ are in correspondence with solutions of the equation
\begin{equation} \label{eqn:flatConnFigure8}
    \lambda^8+\lambda^{-8} -2=\lambda^4+\lambda^{-4}+\lambda+\lambda^{-1}, \qquad \lambda \ne -1
\end{equation}
over $\C$.
This is just the request that the $A$-polynomial~\cite{cooper1998representation} of the figure-eight knot,
\begin{equation}
    A(\lambda,\mu) =  - 2 +\mu^4+\mu^{-4}-\mu^2-\mu^{-2} - \lambda - \lambda^{-1},
\end{equation}
vanishes when we enforce the gluing condition $\lambda^2 = \mu$. Note that if $\lambda$ solves the equation, then also $\lambda^{-1}$ does. Moreover, we have to discard the solution $\lambda = -1$, which has multiplicity $2$, leaving us with a degree 14 polynomial equation. 
Thus, we find $14$ solutions in $7$ pairs $(\lambda,\lambda^{-1})$, since $\C$ is algebraically closed, which correspond to the $7$ isolated points of the moduli space of non-trivial $\SL(2,\C)$ flat connections $\mathcal{M}^*(S_{-1/2}(4_1),\SL(2,\C))$. These are well-known and already appeared in the literature, see for example~\cite{wheeler2025quantum}. The values of the Chern-Simons action $\CS[\gamma]$ are listed in Table~\ref{tab:CSvaluesFigure8}.
Notice in particular that $\CS[\gamma_6]$ and $\CS[\gamma_7]$ are complex. 
\begin{table}[ht!]
    \centering
    \begin{tabular}{Sc | Sc | Sc| Sc | Sc| Sc | Sc| Sc}
         Flat connection & $\gamma_1$ & $\gamma_2$ & $\gamma_3$ & $\gamma_4$ & $\gamma_5$ & $\gamma_6$ & $\gamma_7$   \\
         \hline
         Approx.\ $\CS[\gamma]$ & $ 0.514$ & $ 0.828$ & $ 0.997$ & $  0.235$ & $ 0.054$ & $ 0.123 - 0.035i$ & $ 0.123 + 0.035i$
    \end{tabular}
    \caption{Approximate values of the Chern-Simons action on the $\SL(2,\C)$ flat connections for $M =S_{-1/2}(4_1)$.}
    \label{tab:CSvaluesFigure8}
\end{table}

\subsubsection*{$\SL(2,\F_q)$ Dijkgraaf-Witten theory}

In looking for all the elements of $\Hom(\pi_1(S_{-1/2}(4_1)),\SL(2,\F_q))$, we remark that it is not enough to consider Equation~\eqref{eqn:flatConnFigure8} over $\F_q$. Indeed, this would find only solutions for which the longitude and meridian are represented by diagonal matrices in $\SL(2,\F_q)$, while we should also take care of the other possible conjugacy classes. The case of codiagonal matrices is easily dealt with by solving Equation~\eqref{eqn:flatConnFigure8} over the quadratic extension $\F = \F_{q^2}$. Finally, one can check that there exist solutions of upper triangular form if and only if $q$ is a power of $p = 7$. We discard such a case as it would violate the assumption that $q$ is generic.
Thus, for a generic $q$, we find at most $7$ non-trivial orbits of $\SL(2,\F_q)$ flat connections, corresponding to pairs of solutions $(\lambda,\lambda^{-1})$ of Equation~\eqref{eqn:flatConnFigure8} over $\F_q$ or $\F_{q^2}$. Each orbit has size equal to $|\SL(2,\F_q)|$.

For completeness, we notice that $p= 1143067$ is the smallest prime for which the polynomial splits over $\F_p$ and all seven flat connections are in diagonal form. We computed the value of the Dijkgraaf-Witten action on these flat connections. Results are listed in Table~\ref{tab:DWvaluesFigure8}. We already noted that our conjecture could not be generalized to hyperbolic 3-manifolds, as Dijkgraaf-Witten theories cannot reproduce complex phases. Furthermore, no evident relation between the $\SL(2,\C)$ Chern-Simons and $\SL(2,\F_q)$ Dijkgraaf-Witten values emerged. Even though the values of the CS and DW invariants do not match, we observe a remarkable correspondence between the leading orders. Indeed, for trivial flat connections $\gamma_0$ and $\beta_0$ we have $\delta_{\gamma_0} = 3/2$ and $\Delta'_{\beta_0}= 3/2$, while $\delta_{\gamma_i} = 0$ and $\Delta'_{\beta_i}=0$ for all the non-trivial connections.
However, this correspondence makes sense only when $q$ is such that Equation~\eqref{eqn:flatConnFigure8} splits over $\F_{q^2}$. Unlike for manifolds with no hyperbolic components, such a condition cannot be realized as a divisibility condition between $q$ and the order of the values of the CS invariant (which might be infinite).

\begin{table}[ht!]
    \centering
    \begin{tabular}{Sc | Sc | Sc| Sc | Sc| Sc | Sc| Sc}
         Flat connection & $\beta_1$ & $\beta_2$ & $\beta_3$ & $\beta_4$ & $\beta_5$ & $\beta_6$ & $\beta_7$   \\
         \hline
         $\DW[\beta]$ & $ \frac{173068}{190511}$ & $ \frac{141395}{190511}$ & $ \frac{109168}{190511}$ & $ \frac{105685}{190511}$ &$ \frac{93571}{190511}$ & $ \frac{20011}{190511}$ & $ \frac{119146}{190511}$
    \end{tabular}
    \caption{Exact values of the Dijkgraaf-Witten action on the $\SL(2,\F_q)$-flat connections for $M =S_{-1/2}(4_1)$ and $q =1143067$.}
    \label{tab:DWvaluesFigure8}
\end{table}

\section{A proof of a weak version for spherical manifolds} \label{sec:sphmflds}

In the previous section, we have seen several examples of spherical manifolds: lens spaces were considered in Section~\ref{sec:lensspaces}
and the Poincaré homology sphere appeared as a special case of a Seifert homology sphere in Section~\ref{sec:seifertHS}. For all of them we managed to check our conjecture as stated in Section~\ref{sec:relationCS}.
However, the proof that the conjecture holds for any 3-manifold is beyond our reach. 

In this section, our goal is to prove for all spherical 3-manifolds a statement that is weaker than~\eqref{eqn:DWconjecture}. It can be stated as follows. 
\begin{theorem} \label{thm:sphericalTheorem}
    There exist infinitely many primes $p$ for which we can find a finite field $\F_q$ of characteristic $p$, and the partition function of $\SL(2,\F_q)$ Dijkgraaf-Witten theory, twisted by the 3-cocycle $\omega$ as in Section~\ref{sec:choiceTwist}, can be expressed as
\begin{equation}
\label{eqn:DWsphericalConjecture}
    Z_\DW(M)=\sum_{Q \in \mathcal{S}_\DW}  e^{2\pi i \, Q} \, T_Q(q),
\end{equation}
where $\mathcal{S}_\DW$ is as in~\eqref{eqn:DWconjecture}, and the $T_Q(q)$ are some rational numbers in $|\SL(2,\F_q)|^{-1} \Z$.
\end{theorem} \noindent
That is, we are only able to prove the result for the phase factor, but not for the leading order in the asymptotic expansion.  Notice that for spherical manifolds there are no additional contributions from $\SL(2,\C)$ flat connections, as $\pi_1(M)$ is finite and the finite order elements of $\SL(2,\C)$ sit in its $\SU(2)$ subgroup. Then, in the notation of Section~\ref{sec:relationCS}, we may assume $\mathcal{S} = \mathcal{S}_{\SU(2)}$.

The main idea of the proof is to apply a result by Nosaka \cite{Nosaka23}: 
\begin{proposition}
    Let $\rho : \SL(2,\F_q) \longto \SL(N,\C)$ be a representation. Call $\mathrm{CCS} \in H^3(\SL(N,\C), \C/\Z)$ the 3-cocycle that computes the CS invariant via pairing with the fundamental class of $M$, i.e.\ $\CS_{\SL(N,\C)}[\gamma] = \langle \gamma^* \mathrm{CCS}_{\SL(N,\C)}, [M]\rangle $. Then
    \begin{equation}
        \rho^*(\mathrm{CCS}_{\SL(N,\C)}) = \beta^{-1}c_2(\rho) \in H^3(\SL(2,\F_q),\Q / \Z),
    \end{equation}
    and the Dijkgraaf-Witten invariant with twist $\beta^{-1}c_2(\rho)$ is given by
    \begin{equation}
        \sum_{f \in \Hom(\pi_1(M), \SL(2,\F_q))} \CS_{\SL(N,\C)}[\rho \circ f].
    \end{equation}
\end{proposition} \noindent
If the representation is $\rho =\Ind_T^G \phi + \Ind_{T'}^G \phi'$, we argue that, when $M$ is spherical, we can replace $\rho \circ f$ with $\text{Sym}^{2q^2-1} \circ \hat f$ for some $\hat f : \pi_1(M) \longto \SL(2,\C)$. Here, $\text{Sym}^N$ is the $N$\textsuperscript{th} symmetric power of the $\SL(2,\C)$-representation given by matrix multiplication. Then, using that
\begin{equation}
    (\text{Sym}^N)^* \mathrm{CCS}_{\SL(N+1,\C)} = \frac{N(N+1)(N+2)}{6} \mathrm{CCS}_{\SL(2,\C)},
\end{equation}
we obtain
\begin{equation}
\label{eqn:pullbackOfCS}
    \DW[f] =  \CS_{\SL(2q^2,\C)}[\rho \circ f] = \frac{q^2 (4q^4-1)}{3} \CS_{\SL(2,\C)}[\hat f],
\end{equation}
where we can interpret the integer prefactor $q^2(4q^4-1)/3$ as the level of the $\SL(2,\C)$ Chern-Simons. In this argument, it looks like we don't need the twelvefold second Chern class of $\rho$, $\omega = 12\beta^{-1}c_2(\rho)$. Indeed, everything works just fine if we use $\beta^{-1}c_2(\rho)$. However, we don't want the value $Q$ in~\eqref{eqn:DWsphericalConjecture} to depend on $q$, i.e.\ we would like the prefactor in~\eqref{eqn:pullbackOfCS} to satisfy
\begin{equation}
\label{eqn:sphericalPrefactor}
    \frac{q^2 (4q^4-1)}{3} Q = Q \mod 1
\end{equation}
for any $Q \in \mathcal{S}$. The infinitely many values of $q$ in Theorem~\ref{thm:sphericalTheorem} have a key property, which we have already encountered. Indeed, we find a correspondence between the orbits  of $\Hom(\pi_1(M),\SL(2,\F_q))$ and $\Hom(\pi_1(M),\SL(2,\C))$ under the respective adjoint actions. For generic $q$, this implies the condition $q^2 = 1 \mod |\pi_1(M)|$. 

For spherical manifolds, it is known\footnote{It suffices to notice that any flat connection $A$ can be lifted to the unique flat connection $A_0$ of $S^3$ via its universal covering $p : S^3 \longto M$. Then, the Chern-Simons invariant of $A_0$ must be equal to the Chern-Simons invariant of $p^*A$. This, on the other hand, is equal to $|\pi_1(M)| \CS[A]$, so that we have obtained $|\pi_1(M)| \CS[A] = 0 \mod 1$. } that the order $o(Q)$ for all the values $Q \in \mathcal{S}$ of the Chern-Simons invariant divides $|\pi_1(M)|$.
Thus we may assume $q^2 = 1 \mod o(Q)$. Notice that this very condition already appeared in the hypothesis of our main conjecture~\eqref{eqn:DWconjecture}. However, this is not enough as one can check that the condition~\eqref{eqn:sphericalPrefactor} is satisfied only if $3 \nmid o(Q)$. Here, using the twelvefold second Chern class of $\rho$ is again crucial. Indeed, everything gets multiplied by $12$ and the condition~\eqref{eqn:sphericalPrefactor} becomes
\begin{equation}
    4 q^2 (4q^4-1) Q = 12 Q \mod 1,
\end{equation}
which is true for any $Q \in \mathcal{S}$.

We now proceed with the proof of the claim $\rho \circ f = \mathrm{Sym}^{2q^2-1} \circ \hat{f}$. For this argument, it is essential that $M$ is a spherical manifold. Recall that a spherical manifold is given as a quotient of $S^3$ by a finite subgroup $\Gamma \subset SO(4)$ and hence, spherical manifolds are classified by their fundamental group $\pi_1(M) = \Gamma$. The possible choices for $\Gamma$ are completely classified, see for example \cite{scott1983geometries}. Such finite groups are either generated by at most two elements, but a third generator may be necessary if they have an extra cyclic factor. This extra generator can only appear in a handful of cases and one can check that it plays no role when we study $\SL(2,\F_q)$ or $\SL(2,\C)$ representations, so it can be forgotten. Thus, we may assume that $\Gamma$ is generated by two elements $s$ and $t$. 

Consider the following diagram:
\[
\begin{tikzcd}[row sep=3.5em, column sep=4.5em]
& \pi_1(M) \arrow[dl, "f"'] \arrow[dr, "\hat{f}"] & \\
\mathrm{SL}(2,\F_q) \arrow[d, "\rho"'] && \mathrm{SL}(2,\mathbb{C}) \arrow[d, "\operatorname{Sym}^{2q^2-1}"] \\
\mathrm{SL}(2q^2,\mathbb{C}) \arrow[rr, dashed, "\sim"] && \mathrm{SL}(2q^2,\mathbb{C})
\end{tikzcd}
\]
Our goal is to find the putative automorphism of $\SL(2q^2, \mathbb{C})$ that makes the diagram commute. 
Notice that the homomorphisms $f$ and $\hat{f}$ are determined by the images of $s$ and $t$, which we call $x,y$ and $a,b$ respectively. Then, we call $X,Y$ and $A,B$ their images in $\SL(2q^2,\mathbb{C})$. To exhibit the automorphism, we will prove that the pairs $(X,Y)$ and $(A,B)$ are simultaneously similar, using a result by Friedland~\cite{friedland1983simultaneous}.
\begin{theorem}
    Two irreducible pairs $(X,Y)$ and $(A,B)$ in $\SL(N,\mathbb{C})$ are simultaneously similar if and only if the polynomials $tr(A^{i_1}B^{j_1} \dots A^{i_m}B^{j_m})$ have the same values on these pairs, for $0\le i_k,j_k \le 1$ and $m = N(N-1)$. 
\end{theorem} \noindent
Notice that we have an additional technical hypothesis about the pairs: they are irreducible. We managed to check this hypothesis in a few explicit cases, but we don't know how to prove it in general. We believe that it should generically hold, up to some sporadic cancellations, due to the properties of $\text{Sym}^N$. To be more explicit, irreducibility then means that $A$ and $B$ do not preserve a common proper linear subspace. Since we can always bring $A$ into diagonal form, it is enough to study subspaces that are left invariant under $B = \text{Sym}^N(b)$. In particular, given $A$ diagonal, the pair $(A,B)$ is irreducible if $B_+ =(|B_{ij}|)$ is such that the matrix $(\I + B_+)^N$ has no vanishing entry. Unless commuting, the relations of $\pi_1(M)$ force all entries of $b$ to be non-zero. Then, even though $\text{Sym}^N(b)$ might have some sporadic vanishing entries, $(A,B)$ is irreducible. However, one should be careful about controlling the cancellations in $B$ to prove the irreducibility rigorously. 

Thus, we have to compute the traces of words in $(X,Y)$ and $(A,B)$ and check whether or not they coincide. 
Since $\Gamma$ is a finite group, any word in the generators will have a finite order $n$. Thus, a word in $X,Y$ will be similar to the image of the representative of some conjugacy class of $\SL(2,\F_q)$. Hence
\begin{equation}
    \text{tr}(W(X,Y)) = 
    \begin{cases}
        \text{tr} ( d(\lambda) ) = \phi(\lambda)+\phi(\lambda^{-1}) ,  \\
        \text{tr} ( d'(\xi) ) = \phi'(\xi)+\phi'(\xi^{-1}), \\
        \text{tr} ( \pm \I ) = \pm 2q^2.  
    \end{cases}
\end{equation}
Notice that it cannot be similar to $\epsilon u_\pm$ since we are away from special primes.

On the other hand, a word in $A,B$ will be similar to the image of some diagonal matrix in $\U(1) \subset \SL(2, \mathbb{C})$. Thus, $\text{tr}(W(A,B)) = U_{2q^2-1}(\text{tr} \ d(\zeta_n)/2)$. After some algebra, we find
\begin{equation}
    \text{tr}(W(A,B)) = 
    \begin{cases}
    \text{tr} (W(a,b)), \qquad &\text{if } n \mid q \pm 1,\\
    \pm (q+1), & \text{if } n = 1,2.
    \end{cases}
\end{equation}
Remarkably, two-generated subgroups of $\SL(2,\C)$ have the property that the trace in any word $W$ of the generators $x$ and $y$ is a polynomial in three variables $P_W(u,v,w)$, where $u = \tr x$, $v = \tr y$ and $w = \tr xy$ \cite{traina1980trace}. This result is a direct consequence of the properties of the trace. Indeed, it can be proven only by using the identities
\begin{align}
    &\tr MN = \tr NM, \\
    &\tr M = \tr M^{-1}, \\
    &\tr M N^{-1} + \tr MN = \tr M \tr N.
\end{align}
Now, it is clear that the same identities apply for $M,N \in \SL(2,\F_q)$. This implies that a similar result holds for two-generated subgroups of $\SL(2,\F_q)$.
Then, taking any word in the generators $s$ and $t$ of $\Gamma$, we have obtained that $\tr W(x,y) = P_W(\tr x, \tr y, \tr xy)$ and $\tr W(a,b) = P_W(\tr a, \tr b, \tr ab)$, where we remark that the two polynomials are exactly the same. Notice this makes sense as $P_W$ is a polynomial with integer coefficients, and hence can be understood as a polynomial over any field. 

This translates to a very strict condition on the homomorphisms $f$ and $\hat f$. Indeed, for any word we can write the equation
\begin{equation}
\label{eqn:characteristicEquations}
    \lambda^2_W - \lambda_W P_W(u,v,w) + 1 = 0,
\end{equation}
where $u = \lambda_s+\lambda_s^{-1}$, $v = \lambda_t+\lambda_t^{-1}$ and $w = \lambda_{st} + \lambda_{st}^{-1}$. It is just the characteristic polynomial equation for a matrix in $\SL(2,\F)$\footnote{If not specified, the field  $\F$ is either $\F = \F_q$ or $\F = \C$.}, having replaced $\tr W $ with $P_W$.
In addition the this set of equations, we have additional constraints that come from the relations of $\Gamma$. In particular, $s$, $t$ and $st$ have a fixed order and this yields
\begin{equation}
\label{eqn:orderEquations}
    \lambda_s^m = \lambda_t^n = \lambda_{st}^l = 1.
\end{equation}
Notice that choosing $\lambda_s$, $\lambda_t$ and $\lambda_{st}$ does not fix uniquely the homomorphism from $\Gamma$ to $\SL(2,\F)$, however it does fix the traces of all possible words.

Now, we want to make use of this set of equations and deduce that when they both have solutions over $\C$ and $\F_q$, then the solutions must be somehow related. This is a well-known result that can be found in e.g.\ \cite{serre2009use}.
\begin{theorem}
    Let $\{ f_1, \dots, f_r\} \subset \Z[x_1, \dots,x_n]$. The set of equations $f_i = 0$ admits a common solution over $\C$ if and only if it admits a common solution over some field of characteristic $p$ for infinitely many primes $p$.
\end{theorem} \noindent
The proof of this statement is based on Hilbert's Nullstellensatz and gives a constructive way to find the finite field solution, given a solution over $\C$. We will now review such construction. 

Suppose that we fix a homomorphism $\hat f:\Gamma \longto \SL(2,\C)$. This produces a solution of the system of equations~\eqref{eqn:characteristicEquations}-\eqref{eqn:orderEquations} that corresponds to taking $\lambda_W$ and $\lambda_W^{-1}$ to be the eigenvalues of the matrix $W(a,b)$. A solution always exists since $W(a,b)$ has finite order and is thus diagonalizable with eigenvalues being roots of unity. Hence, the solution over $\C$ is given in terms of algebraic integers.
Let $E$ be the number field generated by the algebraic integers $\lambda_W$. Call $\mathcal{O}_E$ the ring of integers of $E$.
Then, choose a prime ideal $\mathfrak{p} \subset \mathcal{O}_E$ over $p$, i.e. $\mathcal{O}_E / \mathfrak{p}$ is isomorphic to a finite field $\F_q$ of characteristic $p$. The solution over $\F_q$ can then be obtained by reduction over $\mathfrak{p}$. In particular, one can check that the reduction of roots of unity is inverse to the map $\phi$, for an appropriate choice of the solution over $\C$. Notice that the two maps being inverse to each other makes sense only for roots of unity $\zeta$ and elements $x \in \F_q$ that have the correct order, i.e. only for those that may appear as the eigenvalues of words in $\SL(2,\F)$, which is actually the only case we care about. 
This concludes the proof, as $\tr W(a,b) = \lambda_W + \lambda_W^{-1} =  \phi(\lambda_W \!\!\!\mod \mathfrak{p}) + \phi(\lambda_W \!\!\!\mod \mathfrak{p})^{-1}$.

To be precise, one needs to be a bit careful since it appears that using the above theorem, we cannot produce solutions in any finite field. In particular, some quadratic equations might not have solutions over $\F_q$, but they may still be relevant. This is the case for matrices of the form $d'(\xi)$ where the eigenvalue $\xi$ belongs to the quadratic extension $\F_{q^2}$ of $\F_q$. Hence, in general one would have to solve the equations over $\F_{q^2}$ in order to obtain the appropriate homomorphism $f : \Gamma \longto \SL(2,\F_q)$ which makes the diagram commute. 

We conclude this section with the obvious question of what fails in this proof when we consider non-spherical geometries. The crucial observation is that a word in $A,B$ may have infinite order, while this is not true for a word in $X,Y$ (as an example, consider the word $t^2st^2sts \in \pi_1(M)$ for the Brieskorn homology sphere $M = \Sigma(2,3,7)$). Then, the eigenvalues of $W(a,b)$ are no longer roots of unity, and we can easily produce counterexamples.


\section*{Acknowledgments}
The authors would like to thank Mohamed Aliouane, Anna Beliakova, Sergei Gukov, Fernando Rodriguez Villegas, Ayush Singh, and Paul Wedrich for fruitful discussions on the matter.
The research of T.N.\ is partially supported by the INFN Iniziativa Specifica GAST and Indam GNFM. The research of J.Q.R.\ is partially supported by the INFN Iniziativa Specifica ST\&FI.


\appendix
\section{Antisymmetric cocycles} \label{app:acocycles}

Before delving into the definition of antisymmetric cocycles, it will be useful to represent group cochains as \textit{homogeneous} cochains. An homogeneous $n$-cochain in $C^n(G,A)$ is a map $F:G^{n+1} \to A$ which transforms under a $G$-action as follows:
\begin{equation}
    (g \cdot F) (g_0, \dots, g_n) = F(g g_0, \dots, g g_n).    
\end{equation}
The differential $\delta: C^n(G,A) \to C^{n+1}(G,A)$ is then defined as 
\begin{equation}
    (\delta F)(g_0, \dots, g_{n+1}) = \sum_{i = 0}^n (-1)^i F(g_0, \dots, \widehat{g_i}, \dots, g_{n+1}),  
\end{equation}
where the hat indicates we are removing the corresponding entry from $F$. We can go back and forth between homogeneous and inhomogeneous cochains (the latter defined as in Section \ref{sec:DWsl2q}) through the relation
\begin{equation}
    f(g_1, g_2, \dots, g_n) = F(1, g_1, g_1g_2, \dots, g_1g_2 \hdots g_n).
\end{equation}
We will use capital letters to denote homogeneous cochains and the corresponding small letters for their inhomogeneous versions.

We now define \textit{antisymmetric} group cochains as follows. An $n$-cochain $F$ is antisymmetric if, for any permutation $\sigma \in S_{n+1}$, we have
\begin{equation}
    F(g_{\sigma(0)},\dots, g_{\sigma(n)}) = \varepsilon(\sigma) F(g_0, \dots, g_n),
\end{equation}
where $\varepsilon(\sigma)$ denotes the sign of $\sigma$. It is straightforward to translate these conditions into their inhomogeneous form. For example, one finds Eq.\ (\ref{eq:cocyclegauge}) in the case $n = 3$.

For an arbitrary (normalized) $n$-cochain $F$, it is not always possible to find an antisymmetric cochain $F'$ cohomologous to it, unless certain divisibility conditions are satisfied. For example, it is well known that any 2-cochain can be made antisymmetric whenever every element in $\Im(F)$ has a half inside $A$ (not necessarily unique), i.e.\ if $\Im(F)$ is 2-divisible inside $A$. If this holds, an explicit expression is given by
\begin{equation}
    f'(g,h) = f(g,h) - \delta\alpha(g,h), \qquad \alpha(g) = \frac{1}{2}f(g,g^{-1}).
\end{equation}
For $n = 3$, we find that the analogous condition on $\Im(F)$ is that it is both $2$, $3$ and $4$-divisible inside $A$. If that is the case, an explicit expression reads
\begin{equation}
    f'(g,h,k) = f(g,h,k) - \delta\alpha(g,h,k), 
\end{equation}
where
\begin{multline} \label{eq:anti3cocycle}
    \alpha(g,h) = \frac{1}{2}f(g,g^{-1},gh)-\frac{1}{3}f(gh,h^{-1},h)+\frac{1}{3}f(h,(gh)^{-1},gh)
    -\frac{1}{6}f((gh)^{-1},gh,h^{-1}) \\
    +\frac{1}{6}f(h^{-1},h,(gh)^{-1}) -\frac{1}{4}f(g,g^{-1},g) + \frac{1}{4}f(h,h^{-1},h) - \frac{1}{4}f(gh,(gh)^{-1},gh).
\end{multline}
To the best of our knowledge, such an expression for the $n=3$ case has not been computed in the existing literature.

Let us now focus on our particular setting, where the cocycle $\omega = 12\beta^{-1}c_2(\rho)$ has image $6 \Z_{q^2-1} \subset \U(1)$, with $q$ a prime power. While the operations $1/2, 1/3$ are always well-defined for $\Im(\omega)$, in general the $1/4$ operation is not. For this reason, we instead switch to the cocycle $\omega = 24\beta^{-1}c_2(\rho)$, for which Eq.\ (\ref{eq:anti3cocycle}) can be readily applied.

\section{Class multiplication coefficients} \label{app:sums}

Given a group $G$ with conjugacy classes $C_1, \dots, C_k$, one can introduce the conjugacy class sums as
\begin{equation}
    \mathcal{C}_i \coloneqq \sum_{g \in C_i} g.
\end{equation}
Then, the class multiplication coefficients $c_{ijk}$ are defined by the formula
\begin{equation}
    \mathcal{C}_i \, \mathcal{C}_j = \sum_k c_{ijk} \, \mathcal{C}_k,
\end{equation}
see~\cite{isaacs1994character} for more details. 
An equivalent characterization is given by
\begin{equation}
\label{eqn:classCoeffDef}
    c_{ijk} = |\{ (g_i,g_j,g_k) \in C_i \times C_j \times C_k : g_ig_jg_k^{-1} = e\}|.
\end{equation}
It is not hard to prove that the class multiplication coefficient can also be expressed in terms of the irreducible characters of the group $G$. This is known as the Frobenius character sum formula:
\begin{equation}
    c_{ijk} = \frac{|C_1||C_2||C_3|}{|G|} \sum_{\chi \in \mathrm{Irr}(G)} \frac{\chi(C_1) \chi(C_2)\chi(C_3)^*}{\chi(e)}.
\end{equation}

The equivalent characterization~\eqref{eqn:classCoeffDef} prompts us to generalize the definition of the class multiplication coefficients to more than three conjugacy classes. Then, the Frobenius formula holds in a similar fashion. 
Indeed, let
\begin{equation}
    \mathcal{N}(C_1, \dots, C_k) \coloneqq |\{ (g_1, \dots, g_k) \in C_1 \times \dots \times C_k : g_1 \cdots g_k = e\} |
\end{equation}
be the generalization of the class multiplication coefficients, where $\mathcal{N}(C_i,C_j,C_k)$ recovers\footnote{Note that one should be careful in comparing $c_{ijk}$ with $\mathcal{N}(C_i,C_j,C_k)$, as we follow a different notation where $g_3^{-1}$ is replaced with $g_3$ for a neater result.} $c_{ijk}$. One can check that the Frobenius formula becomes
\begin{equation}
\label{eqn:generalizedFrobenius}
         \mathcal{N}(C_1, \dots, C_k) = \frac{|C_1|\cdots|C_k|}{|G|} \sum_{\chi \in \mathrm{Irr}(G)} \frac{\chi(C_1) \cdots\chi(C_k)}{\chi(e)^{k-2}}.
\end{equation}

Notice that expressions of this form appear in the DW partition function of plumbed manifolds at any vertex of degree at least three. In particular, this is the case for the partition function of Brieskorn homology spheres considered in Section~\ref{sec:seifertHS}. Consider $a_i$ be as in~\eqref{eqn:phsPFpart1}, we have:
\begin{equation}
    \frac{|\SL(2,\F_q)|}{q_{\pm}^3} \sum_{\chi} \frac{\chi(a_1) \chi(a_2) \chi(a_3) \chi(-\I)}{\chi(\I)^2} =  \frac{1}{|\SL(2,\F_q)|} \mathcal{N}(a_1,a_2,a_3,-\I).
\end{equation}

The computation of $\mathcal{N}(a_1,a_2,a_3,-\I)$ is highly non-trivial. One way to do it is based on the same approach that appeared in~\cite{andersen2022resurgence} to find elements of $\Hom(\pi_1(M),\SL(2,\C))$. Then, applying the same idea to $\SL(2,\C)$, we can prove that
\begin{equation}
     \mathcal{N}(a_1,a_2,a_3,-\I) = |\SL(2,\F_q)|.
\end{equation}
A different approach is possible and would consist in evaluating the sum~\eqref{eqn:generalizedFrobenius} directly using the character table of $\SL(2,q)$ (see Table \ref{tab:sl2qchars}). This procedure is more straightforward but is harder to generalize to an arbitrary number of conjugacy classes $k$.


\bibliographystyle{JHEP}
\bibliography{refs.bib}

\end{document}